\documentclass[sigplan,10pt]{acmart}
\pdfoutput=1 %

\makeatletter %
\def\mdseries@tt{m}
\makeatother

\renewcommand\footnotetextcopyrightpermission[1]{} %
\acmConference[CVE-2017-5753 Advisory]{}{July 10, 2018}{}

\acmYear{2018}
\acmISBN{} %
\acmDOI{} %
\startPage{1}

\setcopyright{none}
\usepackage{booktabs}   %
\usepackage{subcaption} %
\usepackage[amssymb]{SIunits}            %
\usepackage{sistyle}
\usepackage{courier}            %
\usepackage{alltt} %
\usepackage{url}                  %
\usepackage{listings}          %

\usepackage[frozencache=true]{minted}          %

\usepackage{enumitem}      %

\usepackage{color} %
\usepackage{graphics}
\usepackage{graphicx}
\usepackage{multirow}
\usepackage{caption}

\usepackage{references}

\usepackage{multirow}
\usepackage{array}
\usepackage{paralist}
\usepackage[xcolor]{changebar}

\begin{document}
\title[Spectre1.1]{Speculative Buffer Overflows: Attacks and Defenses
}         %
\author{Vladimir Kiriansky}
\affiliation{
  \institution{}            %
}
\email{vlk@csail.mit.edu}          %

\author{Carl Waldspurger}
\affiliation{
  \institution{}            %
}
\email{carl@waldspurger.org}          %

\affiliation{~~~}
\affiliation{~~~}

\newcommand{\redline}[1]{%
  \cbcolor{red}
  \begin{changebar}
    #1
  \end{changebar}%
  }%

\newcommand\Sloth{SLoth}
\newcommand\SlothBear{SLoth Bear}
\newcommand\ArcticSloth{Arctic SLoth}

\newcommand{\mytimes}{$\times$}

\newcommand\naive{na\"ive{ }}

\newcommand{\floor}[1]          {\left\lfloor #1 \right\rfloor}
\newcommand{\ceil}[1]           {\left\lceil #1 \right\rceil}
\newcommand{\tuple}[1]            {\ifmmode{\left\langle #1 \right\rangle}
                                 \else{$\left\langle${#1}$\right\rangle$}\fi}
\newcommand{\spectre}[1]{\textsc{Spectre{#1}}}

\definecolor{linenumcolor}{rgb}{0.5,0,0.5} %
\definecolor{myschedule}{rgb}{0.858, 0.188, 0.478}

\setminted[c]{xleftmargin=16pt}
\setminted[asm]{xleftmargin=16pt}

\renewcommand{\theFancyVerbLine}{\sffamily
  \textcolor{linenumcolor}{\scriptsize
    \oldstylenums{\arabic{FancyVerbLine}}}}

\begin{abstract}

Practical attacks that exploit speculative execution can leak
confidential information via microarchitectural side channels.  The
recently-demonstrated Spectre attacks leverage
speculative loads which circumvent access checks to read
memory-resident secrets, transmitting them to an attacker using
cache timing or other covert communication channels.

We introduce \spectre{1.1}, a new Spectre-v1 variant that
leverages speculative {\em stores} to create {\em speculative buffer
overflows}.  Much like classic buffer overflows, speculative
out-of-bounds stores can modify data and code pointers.
Data-value attacks can bypass some Spectre-v1
mitigations, either directly or by redirecting control flow.
Control-flow attacks enable arbitrary speculative
code execution, which can bypass fence instructions and all
other software mitigations for previous speculative-execution attacks.
It is easy to construct return-oriented-programming (ROP) gadgets
that can be used to build alternative attack payloads.

We also present \spectre{1.2}: on CPUs that do not enforce read/write
protections, speculative stores can overwrite {\em read-only}
data and code pointers to breach sandboxes.

We highlight new risks posed by these vulnerabilities, discuss possible
software mitigations, and sketch microarchitectural mechanisms
that could serve as hardware defenses.
  We have not yet
  evaluated the performance impact of our proposed software
  and hardware mitigations.  We describe the salient vulnerability
  features and additional hypothetical attack scenarios only
  to the detail necessary to guide hardware and software vendors in
  threat analysis and mitigations.  We advise users to
  refer to more user-friendly vendor recommendations for
  mitigations against speculative
  buffer overflows or available patches.

\end{abstract}

\maketitle

\section{Introduction}
\label{sec:intro}

We dub the primary new attack mechanism described in this paper
\spectre{1.1} (CVE-2018-3693, bounds check bypass on stores), to
distinguish it from the original speculative execution attack variant
1 (CVE-2017-5753), which we refer to as \spectre{1.0}.  We consider
\spectre{1.1} a minor variant in the variant 1 family, since it uses
the same opening in the speculative execution window --- conditional
branch speculation.

\setcounter{subsection}{-1}

\subsection{\spectre{1.0}: Bounds Check Bypass on Loads}

Allowing execution past conditional branches is the most important
performance optimization employed by speculative out-of-order
processors --- essentially every modern high-performance CPU.  Recently,
multiple independent researchers have disclosed ways for attackers to
leak sensitive data across trust boundaries by exploiting
speculative execution~\cite{horn18spectre,Kocher2018spectre,Lipp2018meltdown}.
Using speculative execution, an attacker is able to influence code in the
victim's domain to access and transmit a chosen secret~\cite{horn18spectre,Kocher2018spectre}.

The transmission channel in current proof-of-concept
attacks uses microarchitectural cache state --- a channel available to
speculatively-executed instructions.  Cache tag state was a
previously-known channel for transmitting information in more limited
scenarios --- side channels (during execution of cryptographic
software operating on a secret~\cite{brumley2005rsa}), and covert
channels (where a cooperating transmitter is used).

The previous \spectre{1.0} attack, as well as currently deployed
mitigations, target sequences like \lstref{spec10}.  Since a
speculative out-of-order processor may ignore the bounds check on line
1, an attacker-controlled value \verb|x| is not constrained by 
\verb|lenb|, the length of array \verb|b|.  A \textit{secret value}
addressable as \verb|b[x]| can therefore be used to influence the index of
a dependent load from array \verb|a| into the cache.

In the simplest attack scenario, the attacker also has access to array
\verb|a|, and flushes it from the cache before executing the
victim code~\cite{yarom2014flush}.  The speculative
attack leaves a footprint in the cache, and the attacker measures each
cache line of \verb|a| to determine which one has the lowest access
time --- inferring the secret value from the address of the fastest
line.  Generic mitigations against \spectre{1.0} and its variants,
such as restricting shared memory or reducing timer precision, 
have been limited to this particular ex-filtration method.

\begin{flushleft}

\begin{listing}[b]
  \centering

\begin{minted}[gobble=4,linenos,escapeinside=~~]{c}
    ~\textbf{if} ~(x < ~{\textit{lenb}}~)
      ~\textbf{return}~ a[~\textcolor{red}{b[x]}~ * 512];
\end{minted}

\caption{\spectre{1.0}: Bounds Check Bypass (\textbf{\em on Loads}).
  Speculative secret access via attacker-controlled {\tt x},
  and indirect-load transmission gadget using attacker-controlled
  cache state for %
  array {\tt a}. %
}
\label{lst:spec10}
\end{listing}
\end{flushleft}

\begin{flushleft}

\begin{listing}[b]
  \centering
\begin{minted}[gobble=4,linenos,firstnumber=last,escapeinside=~~]{c}
    ~\textbf{if} ~(y < ~{\textit{lenc}}~)
      ~\textcolor{red}{c[y]}~ = ~\textcolor{blue}z~;
\end{minted}
\caption{\spectre{1.1}: Bounds Check Bypass (\textbf{\em on Stores}).
  Arbitrary speculative write
  with attacker-controlled {\tt y}, and attacker-controlled or
  known value {\tt z}. %
}
\label{lst:spec11}
\end{listing}
\end{flushleft}

\subsection{\spectre{1.1}: Bounds Check Bypass on Stores}

Code vulnerable to \spectre{1.1} is shown in \lstref{spec11}. 
During speculative execution, the processor may ignore the bounds
check on line 3.  This provides an 
attacker with the full power of an arbitrary write.
While this is only a {\em speculative} write, which leaves no
architecturally-visible effects, it can still lead to information
disclosure via side channels.

As a simple proof-of-concept attack, suppose \verb|c[y]| points to the
return address on the stack, and \verb|z| contains the address of line 2 
in \lstref{spec10}.  During speculative execution of a function return,
execution will be resteered to the transmission gadget, as previously
described.  Note that even if a fence instruction
(e.g., {\tt lfence} or {\tt csdb}~\cite{grisenthwaite18v1}) is
added between lines 1 and 2 to mitigate against \spectre{1.0}, an
attacker can simply adjust \verb|z| to ``jump over the fence''.
Return-oriented-programming (ROP) techniques can also be
used to build alternative attack payloads, as described in
\secref{design}.

In a speculative data attack, an attacker can (temporarily) overwrite
data used by a subsequent \spectre{1.0} gadget.  Performant
gadget mitigations use data-dependent truncation
(e.g., {\tt x \verb|&=| (lenb-1)}) rather than fences.
An attacker regains arbitrary read access by overwriting either the
base of array {\tt b} (line 2), or its length, {\tt lenb} (line 1).

\subsection{\spectre{1.2}: Read-only Protection Bypass}

\spectre{3.0}, aka Meltdown~\cite{Lipp2018meltdown}, relies on lazy
enforcement of User/Supervisor protection flags for page-table entries
(PTEs).  The same mechanism can also be used to bypass the Read/Write
PTE flags.  We introduce \spectre{1.2}, a minor variant of
Spectre-v1 which depends on lazy PTE enforcement, similar to Spectre-v3.
In a \spectre{1.2} attack, speculative
stores are allowed to overwrite {\em read-only} data, code pointers, and
code metadata, including vtables, GOT/IAT, and control-flow mitigation
metadata.  As a result, sandboxing that depends on hardware 
enforcement of read-only memory is rendered ineffective.

\subsection{Current Software Defenses}

Currently, no effective static analysis or compiler instrumentation is
available to generically detect or mitigate \spectre{1.1}.  Manual
mitigations for \spectre{1.0} target only very specific cases in
trusted code (e.g., in the Linux kernel), where a
load is used for further indirect memory accesses.

While generic mitigations for \spectre{1.0} have been productized,
such as compiler analysis for C~\cite{msvc-spectre-variant-1}, they
identify only a subset of vulnerable indirect-load code instances. A
comprehensive compiler-based mitigation approach using speculative
load hardening~\cite{llvm-spectre-variant-1} has been proposed, but
incurs a high performance cost.  Generic mitigations for \spectre{1.0}
deployed for JavaScript, as in V8~\cite{google2018v8-65} and
Chakra~\cite{microsoft-chakra-spectre}, protect only bounds checks for
loads.

If we must rely on software mitigations that require developers to
manually reason about the necessity of mitigations, we may face
decades of speculative-execution attacks.  The limited success at
educating software developers for the past thirty years since the 1988
public demonstration of classic buffer overflows is a cautionary
guide.  The silver lining is that the same coding patterns are vulnerable to
speculative buffer overflows.  A good first step toward
preventing them would be to strengthen existing
checks against stack overflows, heap overflows, integer overflows, etc.

\subsection{Contributions and Organization}

We make several key contributions:
\begin{compactitem}

  \item We introduce \textit{speculative buffer overflows} --- attacks
    based on speculative stores that break type and memory safety
    during speculative execution.

  \item We analyze salient hardware features to guide possible
    software and hardware mitigations.
  \item We present new risks posed by {\em impossible paths},
    {\em ghosts}, and {\em halos}, and discuss possible defenses.

  \item We propose the {\em \Sloth{}} family of microarchitectural
    mechanisms to defend against speculative buffer overflow attacks
    by reducing speculative store-to-load forwarding opportunities for
    attackers.
  \item We present a preliminary threat analysis that indicates
    attackers may be able to mount both local and remote
    confidentiality, integrity, and availability attacks.

\end{compactitem}

In the next section, we provide relevant hardware and software
background related to speculative execution. \secref{analysis}
presents a detailed analysis of speculative buffer overflows,
including both vulnerability mechanisms and possible software
mitigations.  We introduce our \Sloth{} family of hardware
mitigations in \secref{zoo}.  \secref{design} focuses on threat
analysis of payloads leading to remote confidentiality attacks
and local integrity attacks. Finally, we summarize our conclusions
and highlight opportunities for future work in \secref{conclusion}.

\section{Hardware and Software Background}
\label{sec:related}

We first review relevant speculative-execution performance
optimizations of modern out-of-order superscalar CPUs in
\secref{sub:specooo}.  We then describe the hardware features salient
to our minor variants in \secref{sub:stlf} for \spectre{1.1}, and
\secref{sub:rdcs} for \spectre{1.2}.  \secref{sub:rob} discusses
further hardware and software features that impact exploitation
success.

\subsection{Speculative Out-Of-Order Execution}
\label{sec:sub:specooo}

Speculative-execution hardware vulnerabilities are the result of
classic computer architecture optimizations from pre-Internet-era design
decisions.  There are three main optimizations that depend on
speculative execution: branch speculation, exception speculation, and
address speculation.  The currently-disclosed
Spectre variants 1 (bounds check bypass) and
2 (branch target injection) use branch speculation,
variant 3 (rogue load) uses exception speculation, and
variant 4 (speculative store bypass) is one case of address speculation.

Branch speculation takes advantage of temporal
and spatial locality in program control flow, and for most programs
achieves low branch misprediction rates; high-performance
microarchitectures speculate through multiple branches.  Exception
speculation assumes that most operations, e.g., loads, do not need to
trap.  Address speculation is used for memory disambiguation, when
loads are assumed not to conflict with earlier stores to unknown
addresses. Loads are also speculated to hit L1 caches, and
immediately-dependent instructions may observe \textit{value speculation}
with the value 0 (before mini-replay~\cite{horn18spectre,yeager96mipsr10k}).
The first two speculation types are control speculations, and all
subsequent instructions are killed; for the third type, only loads and
their dependent instructions need to be replayed.  

Attempts to expose all three major speculation mechanisms to software
--- respectively, via predication, speculative loads, and advanced loads
~\cite{cnet2005itanium} --- have been largely unsuccessful.  Modern
instruction set architectures (ISAs), such as
RISC-V~\cite{waterman2014riscv} and ARMv8~\cite{arm15trm-cortexA72}, are
designed to assume high-performance CPUs will use speculation
techniques implemented in out-of-order hardware.  As a result,
they avoid introducing features such as branch hints and
predicated execution, and specify a relaxed memory-ordering model.

\subsection{Speculative Store-to-Load Forwarding}
\label{sec:sub:stlf}

The distinctive feature of \spectre{1.1} is its dependence on a
performance optimization that is usually called \textit{store-to-load
  forwarding}.  
A \textit{store buffer} is a microarchitectural structure that tracks stores from instruction issue until they are written back to data caches.
On modern cores, such as Intel's Skylake~\cite{intel17opt}, which tracks
up to 56 in-flight stores, it serves a quadruple duty.

First, as for in-order cores, the store buffer
serves as a write buffer to the L1 cache.  Second, on
out-of-order cores, speculatively-executed stores are never written
back until they retire, i.e., become ``senior stores''.  Third, the
store address and the store data are executed out-of-order as separate
micro-ops, which is useful when addresses are known much earlier than
data.  Fourth, a store buffer is used to ensure memory consistency and
coherence, i.e., processors observe their own stores and stores from
other SMP CPUs.  Memory ordering models for most current ISAs specify that a
load following a store with a matching address observes the stored value,
requiring non-speculative store-to-load forwarding.

\textit{Speculative} store-to-load forwarding is therefore an
optimization that allows a load to execute speculatively using prior
store data as soon as both the store address and data are available.
The requirements are that the load size is no larger than the store
size~\cite{intel17opt}, and the store is the youngest at that address.
The load and store physical addresses must be fully matched;
address speculation techniques which use virtual addresses~\cite{boggs04itj-prescott}
or partial physical tags would be subject to (hypothetical) aliasing attacks.

\subsection{Data TLB Speculation}
\label{sec:sub:rdcs}

Deferring the handling of data TLB page faults until a load commits is an
exception-speculation mechanism used to deliver precise exceptions.
\spectre{3.0} (Meltdown) affects CPUs that do not nullify values on
exceptions, e.g., Intel, ARM, and IBM, but not AMD.  In our 
taxonomy, we use \spectre{3.1} to refer to Spectre variant 3a, which is a
low-priority vulnerability, adding to the long list of known
bypasses to Kernel ASLR, which we revisit in \secref{sub:classicbo}.

Fortunately, an effective workaround for \spectre{3.0} is to use separate
user and supervisor page tables, e.g., kernel page table
isolation~\cite{kaiser}.  Future Intel processors also plan to feature
Rogue Data Cache Load (\verb|RDCL_NO|) protection~\cite{intel18spec-v4}.
However, these approaches do not address \spectre{1.2}.

\paragraph{\spectre{1.2}: Speculative Store Read-only Overwrite}

We have validated this attack on both ARM and Intel x86 processors.
We hope that a Rogue Data Cache \textit{Store}
protection feature can be included in future Intel processors to defend
against our \spectre{1.2} variant.
Ideally, speculative store data should not
be forwarded to dependent loads until the TLB entries have been checked to confirm write privileges.  Alternatively, only the value 0
should be forwarded on a fault, which is safe as long as partial store-to-load
forwarding is not allowed, as noted in \secref{sub:stlf}.

\subsection{Speculative Execution Window}
\label{sec:sub:rob}

There are two main limits for speculative attack execution --- the
maximum number of speculative instructions in flight, and the maximum
delay of branch resolution (in both cycles and instructions).
Current processors support large speculative windows.  For example,
the re-order buffer (ROB) on Intel's Skylake has space for 224
micro-ops, or about 200 instructions for typical code.  Each SMT
thread is allotted half, so an attack must complete within
roughly 100 instructions.

A DRAM reference on a modern server can take 80--200~ns
(60--100~ns on desktops). At a typical clock frequency of 2.5GHz, with
the average instructions per cycle for systems code (IPC 1), and
typical micro-ops per instruction (UPI 1.1), waiting on one DRAM
reference can fill the entire window.

In addition to opening the speculative execution window, an attack is
possible only until the window closes --- when a branch is
resolved and wrong-path instructions are flushed, or when an explicit
fence is reached. Even if attackers do not have any influence over
branch history, an attack opportunity is presented by sensitive branch
mispredictions --- when a branch is taken when it should not have been, or
when a branch is not taken when it should have been.

\paragraph{Superscalar Execution}

A modern superscalar core can execute up to 8 speculative
micro-ops in a given cycle (and up to 4 instructions can commit non-speculative results).
For example, in the same cycle Intel's Skylake can execute up to four
arithmetic instructions or up to two branches, as well as two loads
and one store.  
For \spectre{1.0}, even if a bounds-checking conditional branch is
resolved quickly, the few instructions needed for an
attack gadget may still execute on a superscalar machine.

\paragraph{Non-blocking Caches}

Helping cores scale the ``memory wall'' is the most compelling reason
for speculative execution, and modern out-of-order CPUs attempt to
uncover independent memory requests.  A \textit{non-blocking cache}
allows memory requests past predicted branches to be processed while
waiting on older instructions.

The state of cache lines with outstanding cache misses is handled
in a small number of Miss Status Holding Registers
(MSHRs)~\cite{kroft81isca-mshr}.  For example, Intel's Haswell
microarchitecture maintains 10 L1 MSHRs (Line Fill Buffers) for
handling outstanding L1 misses~\cite{intel17opt}.  Similarly, on the
high-performance ARM A72 processor, 6 L1 MSHRs support up to six
unique cache lines targeted by outstanding
cache misses~\cite{arm15trm-cortexA72}.

Since speculative memory requests that have missed in the L1 cache are not
canceled, initiating a request and placing it in an MSHR
within the speculative execution window is sufficient for a load to be cached.  An
attacker may simply repeat multiple re-executions in order to use
values cached after previous attempts.  For example, consider
{\tt a[b[i]*512]} --- a typical ex-filtration gadget that uses an indirect
load to form a cache side-channel transmitter. The first attempt
ensures the secret value {\tt secret=b[i]} is cached, and subsequent
attempts will refer to that value to compute the indirect address and
reference {\tt a[secret*512]}.

\section{Speculative Buffer Overflows}
\label{sec:analysis}

Speculative buffer overflows allow attackers to execute arbitrary
untrusted code within the victim domain.  To help explain
the hardware mechanisms involved, we dissect our demonstration from C
to assembly to RISC micro-ops (in \secref{sub:asmuops}), and discuss 
longer speculative window requirements (in \secref{sub:precond}).

We elaborate on manual mitigations in \secref{sub:manual}.
\secref{sub:classicbo} considers classic buffer overflow mitigations,
and discusses our proposals for repurposing them to protect against
speculative buffer overflows.

\subsection{\spectre{1.1} Assembly and Micro-ops}
\label{sec:sub:asmuops}

We have validated this attack on both ARM and Intel x86
processors\footnote{Hypothesized on January 19, 2018; tested on
  February 11, 2018; reported on February 12, 2018.
  In addition to Intel and ARM, we also
  provided proof-of-concept code for both \spectre{1.1} and \spectre{1.2}
  to AMD, Google, IBM, and Microsoft.}, but we limit our
exposition to x86-64 assembly.

\begin{flushleft}

\begin{listing}
  \centering
\begin{minted}[gobble=4,linenos,escapeinside=~~]{c}

~\textcolor{black}{void}~ f(u64 x, u64 y, u64 z) {
    ~\textbf{if} ~(y < ~{\textit{lenc}}~)
      ~\textcolor{red}{c[y]}~ = ~\textcolor{blue}z~;
}
\end{minted}
\caption{\spectre{1.1} Vulnerable Function.  On 64-bit processors, a
  64-bit write must be used to overwrite code-pointers. }
\label{lst:spec11_full}
\end{listing}
\end{flushleft}

\begin{flushleft}

\begin{listing}
  \centering
\begin{minted}[gobble=4,linenos,firstnumber=last,escapeinside=~~]{asm}
   cmp    %
   jbe    1f ; taken
~1:~ retq
   ... caller
\end{minted}
\caption{\spectre{1.1}: Retired Instructions (x86-64).  Correct path after attack. }
\label{lst:spec11_retired}
\end{listing}
\end{flushleft}

\begin{flushleft}

\begin{listing}
  \centering
\begin{minted}[gobble=4,linenos,escapeinside=~~]{asm}
   cmp    %
   jbe    1f ; predicted not taken
   mov    c, %
   mov    %
~1:~ retq
   ... caller
\end{minted}
\caption{\spectre{1.1}: Speculated Instructions (x86-64).  Speculated path before attack; RSB predicts return target correctly.}
\label{lst:spec11_before}
\end{listing}
\end{flushleft}

\begin{flushleft}

\begin{listing}
  \centering
\begin{minted}[gobble=4,linenos,escapeinside=~~]{asm}
   cmp    %
   jbe    1f         ; unresolved
   mov    c, %
   mov    %
   ; overwrites (%
~1:~ retq  ; store-to-load forwarding
   ... ~\textcolor{red}{ROP gadget}~
\end{minted}
\caption{\spectre{1.1}: Speculated Instructions (x86-64).  Speculated path during attack; execution resteered.}
\label{lst:spec11_during}
\end{listing}
\end{flushleft}

Line 2 of the C code for \lstref{spec11_full} compiles into
the x86-64 assembly code on lines 5 and 6 in \lstref{spec11_retired}.
When the comparison on line 5 depends on a non-cached data
value, the branch on line 6 (in \lstref{spec11_retired} showing the
correct path) is slow to resolve, which opens a large speculative-execution
window.  \lstreftwo{spec11_before}{spec11_during} show the
active speculative paths before and during an attack.

\lstref{spec11_uop} breaks down the last two instructions from
\lstref{spec11_during} into RISC micro-ops.  The return instruction
\verb|retq| is internally broken into an \verb|LDA| micro-op (line 10,
\lstref{spec11_uop}) that loads the return address, and an indirect
branch to the loaded value (line 11).  On some CPUs, \verb|LDA| may
additionally execute before the store address is known (line 9).  When
executed using the data from the speculative store (line 8) after
store-to-load forwarding, the \verb|retq| will consider the
Return Stack Buffer (RSB)
prediction to be incorrect, even though it is normally nearly perfect.
Resteered away from the correct caller, the CPU front-end fetches the
ROP gadget.

\begin{flushleft}

\begin{listing}
  \centering
\begin{minted}[gobble=4,linenos,firstnumber=last,escapeinside=~~]{asm}
STD %
STA (%
LDA nip, (%
JR  nip          ~\,~; resteered to ROP
\end{minted}
\caption{\spectre{1.1} in plausible RISC $\micro$ops for x86-64.}
\label{lst:spec11_uop}
\end{listing}
\end{flushleft}

\subsection{Spectre1.1 Attack Preconditions}
\label{sec:sub:precond}

The most vulnerable branches depend on the value of previous long-latency
operations, such as one or more dependent non-cached
memory references, as in \verb|array->length|.

For a \spectre{1.1} code-pointer attack, the speculative window must fit
not only the payload gadget(s), but also all instructions between the
vulnerable conditional branch and the attacked indirect branch, typically
a \verb|ret| instruction. 
Since the \verb|ret| would normally be predicted correctly, the attack
{\em must} speculatively execute this indirect branch using corrupt
data, while a prior conditional branch remains unresolved.
Increasingly, indirect control transfers on x86 use a
\verb|ret|, whether as usual for function return, or
for a \textit{retpolined} indirect call/jump, since retpolines are the
recommended approach for \spectre{2} protection~\cite{retpoline}.

For a \spectre{1.1} data attack, the speculative window must stay open
until after a target \spectre{1.0} sequence is reached normally.
Mitigations against \spectre{1.0} that use a speculation barrier
(e.g., {\tt lfence}) would be effective against a \spectre{1.1}
data attack.  However, most deployed mitigations employ a more efficient
data-dependent sequence, as discussed in \secref{sub:masking}.  In
such cases, a data attack can simply overwrite either the array base
or length.

\subsection{\spectre{1.1} Manual Defenses}
\label{sec:sub:manual}

The software mitigations for preventing an out-of-bounds store for
\spectre{1.1} are similar to mitigations for \spectre{1.0}.  Manual
placement of these mitigations, however, requires analyzing many more
potentially-vulnerable locations in order to achieve security with
good performance.

\subsubsection{Speculation Fences}
A fence incurs a high performance penalty from stopping speculative
execution, but can be added even in cases where bounds are not known.
To ensure that \spectre{1.0} loads are ordered after prior
branches, CPU vendors have updated documentation for
existing fence instructions (e.g., {\tt lfence} on x86),
and added new instructions (e.g., {\tt csdb} on ARM).
Such fences can be used to implement load-speculation barriers:

\begin{minted}[gobble=4,linenos,escapeinside=~~]{c}
 if (x < lenb) {
    load_barrier_nospec();
    return a[b[x]*512];
 }
\end{minted}

Although the x86 {\tt lfence} instruction was originally defined
architecturally as a {\em load fence}, Intel and AMD have clarified
that it serves as a general serializing {\em instruction fence}.
Thus, an {\tt lfence} ensures that no later instruction
will execute, even speculatively,
until all prior instructions have
completed~\cite{intel-bcb-whitepaper,amd-managing-spec}.
Other processor vendors should confirm that stores in particular, or
simply all instructions, are ordered by existing or new fences, to ensure
prior branches are resolved before a store:

\begin{minted}[gobble=4,linenos,escapeinside=~~]{c}
 if (y < lenc) {
    store_barrier_nospec();
    c[y] = z;
 }
\end{minted}

While such fences can be added before potentially-vulnerable stores,
there is a high performance cost to unaffected paths:
\begin{minted}[gobble=4,linenos,escapeinside=~~]{c}
memcpy(~\textcolor{black}{void}~* ~\textcolor{red}{d}~, ~\textcolor{black}{void}~* ~\textcolor{blue}{s}~, ~\textcolor{black}{size\_t}~ n) {
  store_barrier_nospec();
  unsafe_memcpy(d, s, n);
}
\end{minted}

\subsubsection{Coarse Masking (Unsafe)}
An index value can be bounded coarsely by
masking it with the next power of two, as implemented in {\tt asm.js/wasm} by
the V8 JavaScript engine~\cite{google2018v8-65}.
This may be acceptable to prevent reaching secrets with out-of-bounds loads:
\begin{minted}[gobble=4,linenos,escapeinside=~~]{c}
 if (x < b.size) {
    x &= b.mask; // next power of 2
    value = b.start[x];
 }
\end{minted}

However, depending on the layout of data in memory, this approach
may fall short of protecting against out-of-bounds stores.
Without accompanying changes to pad memory regions to exact powers of
two (at the attendant overhead of internal memory fragmentation),
vulnerable locations that break type safety may be reachable.

\subsubsection{Data-dependent Exact Masking}
\label{sec:sub:masking}

Whenever the branch and the potentially-vulnerable store are in the
same function, the most performant solution is to ensure that indices
or pointers are truncated via data-dependent operations.
For example, conditional masking for JavaScript loads in V8
emits code equivalent to:
\begin{minted}[gobble=4,linenos,escapeinside=!!]{c}
 if (x < b.size) {
    // unsafe, use assembly!
    x &= (x < b.size) !\textcolor{red}{?}! ~0UL : 0;
    value = b.start[x];
 }
\end{minted}

Similar sequences must be used to protect stores as well.
The illustrative C code is, however, unsafe.  Safe index truncation
requires equivalent {\tt asm volatile} assembly sequences.
Most ISAs support conditional move instructions that compilers can
emit for the ternary select operator.  The mask selection (line 3)
needs to use a conditional move, but the compiler may convert it to an
unsafe branch instead.  However, this depends on the compiler and
optimization level, as well as any profiled-guided optimizations.

Compilers may also optimize out ``unnecessary'' code based on
assumptions about correct paths, such as those involving congruent branches 
or identical code.  Unfortunately, speculative execution invalidates such
optimizations.

The Linux kernel defines an {\tt array\_index\_nospec()}
macro used to safely mask an index and block
speculation~\cite{linux-doc-speculation}.  On x86
it succinctly uses the subtract-with-borrow {\tt sbb} assembly
instruction (while {\tt sbb} is a vestigial low-throughput instruction,
CPU vendors should offer superscalar versions for future silicon).
On ARM it also includes the necessary {\tt csdb} fence.
We recommend that compiler writers provide a built-in function that
safely performs the operation of truncating an index to zero on overflow.

\subsubsection{Congruent Branch TOCTOU (Hypothetical)}
\label{sec:sub:ghosts}

When a mitigation check and its uses appear in separate basic
blocks, placing checks safely becomes more difficult than
simply strengthening existing ones.  Programmers and compilers
typically assume that branches testing the same conditions, as in {\tt
  if} or {\tt for} statements, behave similarly.

\paragraph{Impossible paths}
Congruent branch pairs are those where,
under correct execution, either both are taken or neither are taken.
Usually these are predicted using global branch history to take
advantage of correlations.  However, this is not guaranteed, and
speculative execution may execute not just \textit{wrong paths}, but
also \textit{impossible paths}.  Invariants about buffer bases or sizes,
index bounds, and loop counts will often be invalidated, e.g.,
due to incorrectly initialized variables in {\tt if/else} branches.
To avoid such time-of-check to time-of-use vulnerabilities, all uses must
have adjacent guards with one of the two recommended mitigations above.

\paragraph{Ghosts}
Short loop trip counts (e.g., under 30 \cite{intel17opt,horn18spectre}) are
predicted perfectly by modern path-dependent branch predictors.
However, attackers may prime these predictors, as well as the
architectural state of stacks or heaps, via prior calls.  An
impossible path can influence uninitialized variables ({\tt if}
statements), and uninitialized or unconstrained values past input vector lengths ({\tt for} loops).
We refer to such pseudo-inputs as \textit{ghosts} and \textit{halos},
respectively.

As illustrated in \lstref{ghost}, ghosts allow arbitrary speculative
reads, writes, and code execution.  Ghosts can be avoided by adding a
fence (line 6), or when possible, by modifying program logic.  Explicit manual initialization
with compiler warning assistance
or automatic
zero-initialization \cite{lu2017ndss-stackspray} would need to use
flow-\textit{insensitive} analyses to avoid optimizing out initialization.

\begin{flushleft}

\begin{listing}
  \centering
\begin{minted}[gobble=4,linenos,escapeinside=~~]{c}
   A* pa;          // uninitialized
   if (cond)
     ~\textit{pa = new A();}~ // skipped
   ...
   if (cond)
     *~\textcolor{red}{pa}~ = b;
\end{minted}
\caption{A \textit{ghost} write (in C++).  Attacker controls {\tt pa}
  (via unchecked stack contents), allowing arbitrary write.}
\label{lst:ghost}
\end{listing}
\end{flushleft}

\paragraph{Halos}
As shown in \lstref{halo}, halos are positions beyond the expected values
in array {\tt b} which should have been validated by a gateway function.
In this example, the size of array {\tt a} may not be available to the
worker function, and a fence may be too expensive to add.
\begin{flushleft}

\begin{listing}
  \centering
\begin{minted}[gobble=4,linenos,escapeinside=~~]{c}
  for (i=0; i < n; i++) {
     ~\textcolor{black}{int}~ pos = b[i]; // n <= i < lenb
     a[~\textcolor{red}{pos}~] = c[i];
  }
\end{minted}
\caption{A \textit{halo} read. Attacker controls {\tt pos}
  (by indexing beyond the active entries of {\tt b}), allowing arbitrary write.}
\label{lst:halo}
\end{listing}
\end{flushleft}

Halos can be handled more efficiently by clamping iterator variables.
For example, on line 2 we can ensure {\tt i} is less than {\tt n},
but {\tt n} may not be the capacity of array {\tt b}, which may 
contain unsanitized values; {\tt n} depends on a slow dereference.
Clamping {\tt pos} to 0, similar to \texttt{array\_index\_nospec()},
should be safe, as long as this does not break any other invariants,
e.g., \verb|capacity(a[b[i]]) > len(c[i])|.
\vspace*{4pt}
Due to impossible paths and the risk of ghosts and halos,
{\em all} functions should be analyzed for vulnerable patterns,
not only gateway functions known to process untrusted inputs.

\subsection{Fortified Classic Buffer Overflow Mitigations}
\label{sec:sub:classicbo}

Several generic mitigations have been proposed to protect against
classic memory-safety bugs.  Mitigations against code-pointer
attacks can be strengthened to protect against speculative execution
attacks as well.  However, data attacks remain an important concern
for speculative buffer overflows.

\paragraph{Generic Code-Pointer Protections}

Robust mitigations for code-pointer attacks have been developed,
based on program shepherding~\cite{kiriansky02sec-shepherding}
and follow-on work~\cite{abadi05ccs-cfi,kiriansky03tm-cfi-pldi-rejection,
kuznetsov14osdi-cpi,tice14sec-cfigcc}.  Such mitigations, 
including the subset productized in Microsoft's Control Flow Guard
(CFG)~\cite{weston16blackhat-cfg}, can be
strengthened to perform all target validation checks without using conditional
branches.  Similar speculation-safe checks can augment memory integrity checkers \cite{akritidis08sp-wit} to use poisoned write pointers. 

Guards implemented by conditional branches can be replaced with guards
using arithmetic sequences.  Such sequences can conditionally create
non-canonical 64-bit virtual addresses that poison indirect transfer or
write addresses.  For example, by conditionally {\tt XOR}ing either 0 or the
MSB bit, an unlikely security violation will result in a general protection fault.
Removing never-taken conditional branches from the
instruction stream avoids pollution of global branch history and
should improve the prediction accuracy of remaining branches.
Although more predictable branch-predictor behavior might benefit
attackers, this change may be an overall win for both security and
performance.

Metadata in CFG is currently protected by read-only
PTEs, which are insufficient due to \spectre{1.2}, and
must be strengthened.  The CFG
reference monitor is indirectly reached via a read-only code pointer, which
is an easy attack target.  If indirection is desirable, compiled
direct calls to a thunk routine that can be patched should be used
instead.  While speculative stores will succeed at overwriting
read-only code, we do not expect their results to be forwarded to
instruction fetch. Such store-to-\textit{ifetch} optimization is unlikely,
yet x86\footnote{Unlike most ISAs, on x86 self-modifying code does not require a barrier.}
CPU vendors should document this explicitly.

Hardware mitigations, like Intel's future
CET~\cite{kiriansky03meng-rio, intel17cet}, should offer generic
protection, provided that security checks are not evadable micro-ops that
depend on control speculation, and PTE-based protections for shadow
stacks are enforced during speculative execution.  Yet, CET is incompatible
with \textit{retpolines}~\cite{llvm-spectre-variant-2,
  retpoline}, since they employ stack smashing as a stronger mitigation
for \spectre{2}.  Indirect Branch Restricted
Speculation (IBRS) leaves open possibilities for internal
interference of indirect branch targets~\cite{Kocher2018spectre}.

\paragraph{Return Protections}
Simple variations of classic stack canary checks
\cite{frantzen01sec-stackghost,kiriansky02sec-shepherding} can be
added to protect speculative return instructions: {\tt XOR}ing into
the return address the difference between a stack canary value and its
expected value (secret).  This mechanism has the advantage of backwards
compatibility with current compiler mitigations, preserving the same stack layout and
return address for backtraces.  Unfortunately, canaries may be elided by compiler optimizations
when writes on all architectural paths are ``proven safe''~\cite{microsoft13eaten-cookie};
such analysis is invalidated on speculative paths.

Return Stack Buffer (RSB) hardware protections
\cite{kiriansky03meng-rio} can be repurposed against non-sequential speculative
return address overwrites.  For example, hardware protections may
disallow RSB mispredictions from being resolved speculatively
(mini-exceptions), or may prevent speculative store forwarding to {\tt
  ret} (i.e., forward only from senior stores).

\paragraph{ASLR}
Address Space Layout Randomization, which has been deployed for user
processes, OS kernels, and hypervisors, is the weakest classic buffer
overflow mitigation.  Nevertheless, it is the only generic mitigation
currently available against speculative buffer overflows, and it
mitigates both code and data attacks.  However, ASLR is rendered
ineffective by both classic information leaks, e.g., as used in
EternalBlue (CVE-2017-0144)~\cite{wiki-eternalblue},
as well as side-channels against branch
history~\cite{evtyushkin16micro-jaslr} or MMU page-table walkers
\cite{gras2017ndss-AnC}.

A small change to the vulnerable statement on line 3 in
\lstref{spec11_full}, to {\tt c[y] \textbf{\textcolor{red}{+=}} z},
allows a relative overwrite of a code-pointer and sidesteps ASLR with
\spectre{1.1}.  Additionally, \spectre{3.0} and \spectre{3.1} (plus
additional \spectre{3.1.x} variants) can be used to bypass Kernel ASLR
(KASLR).

\paragraph{Memory Protection Keys}

In some ISAs,
applications may attempt to keep sensitive data accessible only under
a \textit{protection key}, such as the Memory Protection Key (MPK)
technology recently added to Intel systems~\cite{intel-sdm-vol3}.
On current hardware, MPK may not be
enforcable due to \spectre{3}, but future processors with hardware
mitigations should be able to prevent \spectre{1.0} gadgets from
accessing secrets.

A \spectre{1.1} speculative code-execution gadget, however, can first
disable these protections before reading a secret.  Significantly more
sophisticated solutions are needed to prevent classic buffer overflows
\cite{vahldiek18corr-erim} from accessing the MPK {\tt wrpkru}
instruction, which modifies protection keys.  For speculative buffer
overflows, however, modifying the architectural behaviour of {\tt
  wrpkru} to internally include {\tt lfence} should prevent misuse
under speculative attacks.

\section{Hardware Mitigations}
\label{sec:zoo}

In this section, we sketch plausible hardware mitigations specific to
\spectre{1.1}. We are also designing more general microarchitectural
support to protect against both known and unknown variants of Spectre,
but this ongoing work is beyond the scope of this paper.

To defend against \spectre{1.1}, we propose the {\em SLoth} family of
microarchitectural mitigations that constrain store-to-load
forwarding. Successive design points have increasing expected
performance, but also increasing hardware and software
complexity: store-to-load blocking (``\SlothBear{}''),
lazy store-to-load forwarding (``\Sloth{}''), and frozen
store-to-load forwarding (``\ArcticSloth{}'').

\subsection{Store-to-Load Blocking}
The ``\SlothBear{}'' mitigation anticipates plausible microcode updates
for existing silicon to prevent store-to-load forwarding either from
speculative stores, or to speculative loads.
The viability of implementing this mitigation in microcode is unknown.
It affects hardware paths similar to \spectre{4}, for which
existing microcode updates to Intel's production silicon offer a backup plan
mitigation (at up to 8\% cost on SPECint~\cite{intel-ucode-spectre4}).
If store-to-load blocking is
possible with minimum complexity, it would provide maximum security
with a minimum trusted computing base (TCB).

This approach is likely to incur high performance overheads, as it may
impact operations such as register spills and C++ member variable accesses.
Nevertheless, this design point would enable a quick response for
unpatched software, while software developers are educated about how
to look for vulnerable code.  Overall, this mitigation offers a good
safety net for users who find its performance acceptable.

\subsection{Lazy Store-to-Load Forwarding}
The ``\Sloth{}'' mitigation uses compiler-marked instructions that are
candidates for forwarding.  For example, compilers may allow
retpolines to smash the stack, and may mark register spills and
restores explicitly.

The low complexity and small TCB of this approach are attractive,
with changes localized to the load-store unit. Performance is
likely to be acceptable even without software changes; with
compiler co-design, it can achieve optimal performance.

\subsection{Frozen Store-to-Load Forwarding}

If error-prone software mitigations are the only practical alternative
solution to this class of speculative execution attacks, a
higher-performance hardware design may be justifiable, despite its
complexity.

The ``\ArcticSloth{}'' mitigation employs dynamic detection of pairs of
stores and loads that are candidates for forwarding.  A simpler
variant can track the load instructions that have previously required store-to-load
forwarding on correct paths, while accepting data from any store.

This requires a stronger hardware address speculation mechanism,
similar to high-performance Alpha processors
\cite{chrysos98isca-store}, which may increase the complexity, power,
and area of current CPUs. Full physical address tags for load and store
instructions would be required to securely track white-listed pairs of
previously-committed instructions.

\section{Speculative Attack Payloads}
\label{sec:design}

Speculative buffer overflows allow arbitrary speculative code
execution within the victim domain.  Yet, these are short code
fragments, limited to roughly a hundred instructions, and have
short-lived ephemeral effects.  In this section, we discuss
hypothetical payloads that attackers can deploy to escape the weak
sandbox of out-of-order execution.

{\em{}The speculative attacks in this section are based on
our hypothetical threat model analysis for \Sloth{}.}
Our {\em preliminary} threat analysis indicates that attackers may be
able to mount {\em both local and remote} confidentiality, integrity,
and availability attacks.  We advise software developers to broaden
the scope of vulnerable software analysis, and system builders to
design generic defense-in-depth mitigations.

\subsection{Threat Model}
\label{sec:sub:threat}

We assume most systems that process untrusted inputs are at risk from
both local and remote attackers.  High-value systems that use or
maintain secret information (user credentials, private keys, etc.) are
the primary concern.

At highest risk are systems that execute untrusted code, including
virtual machines, containers, and sandboxed web browser environments.
Prior threat analysis of microarchitectural in-filtration and
side-channel ex-filtration limited the threat surface to local
information disclosure.  Remote confidentiality attack targets may
also include login, database, and web servers, SSL-terminating
firewalls, etc.

We focus our discussion on \spectre{1.x}, where speculative window
in-filtration is possible based solely on untrusted inputs.  We assume
victims process attacker requests and may respond to
them. (\spectre{2} also allowed in-filtration into instances that do
not communicate with the attacker, a threat only if the attacker and
victim share a core.)

\subsection{Speculative Attack Ingredients}
\label{sec:sub:attacks}

A speculative attack combines several ingredients:

\begin{compactitem}
\item \textit{vulnerable code} -- reachable by unprivileged attackers.
\item \textit{vulnerable data} -- untrusted input, used to trigger an out-of-bound access.
\item \textit{sensitive data} -- known and addressable chosen secrets.
\item \textit{speculative payload data} -- passed addresses (sensitive data and/or channel parameters).
\item \textit{speculative payload code} -- present executable gadgets.
\end{compactitem}

An attacker must be able to reach code susceptible to a hardware
speculation vulnerability that will not be resolved quickly.  The
speculative payload parameters and code must also be under attacker
control.

\paragraph{Exposed Hardware Vulnerability}
The vulnerable code may be affected by its use of a speculative read
or write.  In addition, reads or writes may use addresses
that differ from the intended addresses (shadows or aliases):
\begin{compactitem}
\item \textit{write} -- \spectre{1.1} (the focus of this paper).
\item \textit{read} -- an out-of-bound function pointer read~\cite{horn18spectre}.
\item \textit{shadow} -- same virtual address used in different address spaces.
\item \textit{alias} -- partial physical-address matching, within or across address spaces.
\end{compactitem}

An out-of-bounds or uninitialized read is an instance of \spectre{1.0},
while an out-of-bounds or uninitialized write is an instance of
\spectre{1.1}.  Shadow and alias address speculation vulnerabilities are
instances of \spectre{4}.

We generally assume a \spectre{1.1} vulnerability leverages existing
code, and may further use ROP gadgets, stack pivoting, etc.  For
completeness, a plausible \spectre{1.0} out-of-bounds function pointer
read~\cite{horn18spectre} also allows arbitrary code execution, in
addition to being treated simply as a form of out-of-bounds
load~\cite{Kocher2018spectre}.  Vtables for ghost objects
(\secref{sub:ghosts}) can be exploited in a similar manner;
i.e., a \spectre{1.0.1} sub-minor variant in the current taxonomy.
\Sloth{} does not protect against \spectre{1.0}.
\subsection{Payload Code -- Confidentiality Attacks}

There are generally two types of side channels where an unintended
shared medium is used for covert or unintended communication.  In a
{\em stateful} channel, receivers use footprint timing
(\secref{sub:footprint}); in a {\em stateless} channel, receivers rely
on throughput timing (\secref{sub:throughput}).

\paragraph{Secret Access}
The attacker selects a secret bit or byte to transmit using an
attacker-controlled parameter, such as {\tt x} in \lstref{spec10}.
The secret may be accessed via a simple absolute address, such as {\tt
  b[x]}.
Similar attacks may involve executing more flexible code
sequences to locate a secret by traversing live application pointers.
An attacker may first need to disable any memory protection keys
(\secref{sub:classicbo}).

\paragraph{Secret Transmission}
Depending on the ex-filtration channel, transmission includes any
necessary bit or byte extraction and shifting, e.g.,
{\tt (b[x] \& 0x1)} \verb|<< 9|.  Data flow from the extracted
secret is then directed to an instruction with a generalized data
dependence: an address-dependent memory access, a control-dependent
instruction selection, or a data-dependent variable-latency operation.

\vspace*{4pt}
A generalized attack schema can be composed of one or more
stages~\cite{kiriansky18crypto-dawg}, whose output impacts
only microarchitectural state.
Attacks may also be composed by combining multiple invocations of
payload stages, where later stages use
microarchitectural state as input, instead of secrets.

\subsection{Receiver (Non-Speculative Code)}
Receiving a secret is generally within the attacker domain, and not
speculative.  Timing an access in a sandboxed or virtualized
environment, however, may be subject to coarsened timer precision,
slowing down a local attack.

\paragraph{Amplifying Timer Precision}

A software mitigation deployed in web browsers is to reduce timer
precision~\cite{performance-now}.  For example, Chrome coarsened
{\tt performance.now()} to use 100~$\micro$s granularity, in order
to prevent accurate measurement of events at time scales that are
orders of magnitude smaller, such as a $\sim$100~ns cache miss to DRAM.

However, this only slows down attacks, without preventing them.
\lstref{jsnow} illustrates a 
simple amplification of a timing attack, by requesting multiple cache
lines that correspond to each measured secret bit.
This approach consumes a larger cache footprint as the amplification
factor grows, which may induce evictions.  Nevertheless, the huge
timing difference between accessing numerous mostly-resident
{\em{}vs.}~non-resident lines still provides a strong signal.

\begin{flushleft}

\begin{listing}[b]
  \centering
\begin{minted}[gobble=4,linenos,escapeinside=~~]{c}
     zero = 0;
     t0 = performance.now();
     for (i = 0; i < 1024; i++)
         zero += a[i][1+zero];
     t1 = performance.now();
     \end{minted}

\caption{Cache Timing Amplification (Hypothetical)}
\label{lst:jsnow}
\end{listing}
\end{flushleft}

As a small pessimization designed to reduce memory-level parallelism,
we also carry a dependence through all accesses --- the value of {\tt
  zero} is always 0, since the contents of array {\tt a} are
zero-filled prior to ex-filtration.

\subsection{Footprint Timing Side Channels}
\label{sec:sub:footprint}

A stateful channel allows time-sharing between transmission and
reception, e.g., the attacker can observe the footprint after the
victim code executes.  Although the most commonly demonstrated attack
mechanism uses timing of cache-line presence state, various
microarchitectural resources can be exploited\cite{ge2018survey}, including:
\begin{compactitem}
\item Cache memory state -- cache lines, cache sets,
  replacement metadata~\cite{kiriansky18crypto-dawg},
  fill buffers, write-back buffers, prefetchers.
\item Branch predictor state -- branch target buffer,
  branch history table, branch history register, RSB.
\item Address translation state -- TLB, PTE cache.
\end{compactitem}

The easiest and most well-studied side channel uses
the particular cache state of individual cache lines.  For example,
in flush+reload~\cite{yarom2014flush} the victim and attacker share a
cache line.  In prime+probe~\cite{liu2015llctiming}, the attacker no longer
needs memory shared with the victim, and instead can use
any congruent cache lines to check if any of them has been evicted
from a shared LLC cache set.
Additional variants include 
evict+reload~\cite{liu2015llctiming} and 
flush+flush~\cite{gruss2016flushflush}.  Cache timing attacks have
even been demonstrated in JavaScript~\cite{oren15ccs-spy}, including 
the ability to bypass ASLR~\cite{gras2017ndss-AnC}.

All footprint attacks can be prevented by carefully partitioning
microarchitectural state.  For example,
DAWG~\cite{kiriansky18crypto-dawg} proposes hardware mitigations that
securely partition all microarchitectural memory structures (set-associative
caches, TLBs, PTE caches, etc.) to protect against both
non-speculative and speculative footprint attacks.  A remote
cache-timing \textit{reflection attack} is also outlined in
\cite{kiriansky18crypto-dawg}.

\subsection{Throughput Timing Side Channels}
\label{sec:sub:throughput}
Throughput or contention timing requires transmission and reception to
be concurrent.  Traditional side-channel attacks use contention on
shared resources as an indicator.  In speculative execution, the
victim may additionally interfere with its own instructions that are
executed non-speculatively.  Examples from classic side-channel
attacks measure contention between the victim and attacker for diverse
shared resources:
\begin{compactitem}
\item Cache resources -- slice, bank.
\item OoO execution resources -- execution ports,
  variable-latency ALUs, banked register file,
  load buffer, store buffer, reorder buffer,
  branch order buffer, reservation station,
  physical register files, free lists, etc.
\item System resources -- DRAM, QPI.
\end{compactitem}

\subsubsection{Reflected Throughput}
We generalize this class of channels as measuring victim system
throughput after any temporary microarchitectural state is influenced
by secret data.  An attacker simply measures performance, such as
the number of executed macro-operations. %

In a traditional
SMT attack, the measured thread may be under attacker control.  This method requires SMT sharing, and is viable for
privilege escalation within a single OS.  It is not feasible for
cross-VM attacks in most cloud settings, which typically avoid
scheduling separate VMs onto hardware threads associated with the same
core.  However, some cloud providers do offer low-cost burstable
``micro'' instances that may allow such SMT sharing.

It is also possible to measure the influence of speculative execution
on an SMT peer within the attacker domain.  This approach is plausible
against public cloud instances which share SMT cores only within a
trusted domain, and requires detection of how connections are mapped
to processing threads.

Finally, the self-interference of the victim thread can simply be measured.
This method is the most general, and would be effective for both
sandbox escapes or remote attacks, as it requires only connection
persistence.

\subsubsection{Local and Remote Channel Modulation}
Any busy resource can be used as an indicator of speculative execution
behavior.  Effects that persist even after speculative instruction
cancellation are the easiest to measure.  We consider several diverse
microarchitectural behaviors that hypothetically may be influenced by a speculative
attacker payload on some systems.

\paragraph{MSHR Modulation}
  Since speculative memory requests are not canceled, and each core
  has a limited number of MSHRs (\secref{sub:rob}), modulating the
  memory level parallelism available to non-speculative execution could
  be measured by its impact on throughput.

\paragraph{Variable-Latency ALU Modulation}
Some data-dependent instructions, such as square-root and division
operations that are not fully pipelined, may exhibit variable latency
that can affect peer threads.  Moreover, the throughput of
non-speculative execution will also be impacted when these
instructions are non-cancellable, or if the instruction window does
not prioritize the oldest instruction~\cite{yeager96mipsr10k}.

\paragraph{AVX2 DVFS Modulation}
Speculative instruction selection over Intel's AVX2 instructions could
be used to modulate the reliability and power-saving features of the power control unit.
For example, if AVX2 instructions
are used they may result in a lower maximum TurboBoost frequency~\cite{intel14haswell-uncore}.
  
\paragraph{RDRAND Contention Modulation}
An attacker could modulate the throughput of Intel's high-quality {\tt
  rdrand} random number generator, which is used non-speculatively by
some SSL implementations for handshakes \cite{freier11rfc-ssl3}.  This
could be prevented by having {\tt rdrand} perform an internal {\tt lfence}.
\vspace*{4pt}

This random sample from across the instruction manual illustrates that
modulation opportunities are pervasive.  As seen with defenses against
classic buffer overflows, detecting all bad behaviors will be harder
than having \Sloth{} prevent attackers from taking over speculative
execution.

\subsection{Integrity Attacks}

In addition to confidentiality attacks, cache eviction can be used for
integrity attacks, both indirectly and directly.  Practical
integrity breaches may simply follow a confidentiality breach in
which an attacker steals access credentials.

Hypothetically~\cite{ccelio18barcelona,spechammer}, speculative
execution can also be used to mount a
RowHammer~\cite{kim2014rowhammer,google2015rowhammer} integrity attack, such as
by leveraging indirect load gadgets as a tool for cache
evictions~\cite{gruss2016js-rowhammer}, or by generalizing network
attacks~\cite{tatar18usenix-throwhammer}.  Since the attacker is
operating within the victim domain, non-speculative
mitigations~\cite{brasser17sec-catt} are not effective.  \Sloth{}
prevents execution of such attacks via \spectre{1.1}.

\section{Conclusions}
\label{sec:conclusion}

We have explored new speculative-execution attacks and defenses,
focusing primarily on the use of speculative stores to create
speculative buffer overflows, which we refer to as \spectre{1.1}.  The
ability to perform arbitrary speculative writes presents significant
new risks, including arbitrary speculative execution.  Unfortunately,
this enables both local and remote attacks, even when \spectre{1.0}
gadgets are not present.  It also allows attackers to bypass
recommended software mitigations for previous speculative-execution
attacks.  
Speculative execution of wrong or \textit{impossible} paths creates
speculative bug class \textit{doppelg{\"a}ngers} to the known classes
of pernicious bugs breaking memory and type
safety~\cite{wiki-memory_safety,cwe-cpp_weaknesses}.
Given the heightened public awareness due to Spectre and related
attacks, there is higher consumer and business acceptance of
previously unthinkable performance overheads for security protections.
We hope this opportunity will be used to raise the bar for strong
generic mitigations against both speculative and classic buffer
overflows, as we have outlined
for both software and hardware (in \secref{sub:classicbo}).

We also believe \spectre{1.1} speculative buffer overflows are
completely addressable by hardware (in \secref{zoo}).
Rather than adding to the classic buffer overflow patch burden, future
systems should be able to close this attack vector completely, with
good performance.

We are confident that future secure hardware and software will be able
to retain the performance benefits of speculative-execution
processors.
We hope to make additional progress in this direction, as we 
continue to explore more general microarchitectural support
and software co-design to protect against both existing and
future Spectre variants.
In the short term, there may be a few rough patches (to be
applied).

\begin{acks}                            %

We are grateful to Joel Emer for his early feedback on this work.
Thanks to Matt Miller for his thorough technical review and helpful discussions.
Jason Brandt, Martin Dixon, David Kaplan, and Paul Kocher
also provided valuable feedback and suggestions.

Thanks to Intel for their partial sponsorship of this research conducted in February 2018.

\vspace*{0.15in}
\noindent Any opinions, findings, conclusions or recommendations expressed
in this material are those of the authors.

\end{acks}

\bibliography{uarch,sgx_references}


\begin{thebibliography}{60}


\ifx \showCODEN    \undefined \def \showCODEN     #1{\unskip}     \fi
\ifx \showDOI      \undefined \def \showDOI       #1{#1}\fi
\ifx \showISBNx    \undefined \def \showISBNx     #1{\unskip}     \fi
\ifx \showISBNxiii \undefined \def \showISBNxiii  #1{\unskip}     \fi
\ifx \showISSN     \undefined \def \showISSN      #1{\unskip}     \fi
\ifx \showLCCN     \undefined \def \showLCCN      #1{\unskip}     \fi
\ifx \shownote     \undefined \def \shownote      #1{#1}          \fi
\ifx \showarticletitle \undefined \def \showarticletitle #1{#1}   \fi
\ifx \showURL      \undefined \def \showURL       {\relax}        \fi
\providecommand\bibfield[2]{#2}
\providecommand\bibinfo[2]{#2}
\providecommand\natexlab[1]{#1}
\providecommand\showeprint[2][]{arXiv:#2}

\bibitem[\protect\citeauthoryear{Abadi, Budiu, Erlingsson, and Ligatti}{Abadi
  et~al\mbox{.}}{2005}]%
        {abadi05ccs-cfi}
\bibfield{author}{\bibinfo{person}{Mart\'{\i}n Abadi}, \bibinfo{person}{Mihai
  Budiu}, \bibinfo{person}{\'{U}lfar Erlingsson}, {and} \bibinfo{person}{Jay
  Ligatti}.} \bibinfo{year}{2005}\natexlab{}.
\newblock \showarticletitle{Control-flow Integrity}. In
  \bibinfo{booktitle}{\emph{Proceedings of the 12th ACM Conference on Computer
  and Communications Security}} \emph{(\bibinfo{series}{CCS '05})}.
  \bibinfo{publisher}{ACM}, \bibinfo{address}{New York, NY, USA},
  \bibinfo{pages}{340--353}.
\newblock
\showISBNx{1-59593-226-7}
\urldef\tempurl%
\url{https://doi.org/10.1145/1102120.1102165}
\showDOI{\tempurl}


\bibitem[\protect\citeauthoryear{{Advanced Micro Devices, Inc.}}{{Advanced
  Micro Devices, Inc.}}{2018}]%
        {amd-managing-spec}
\bibfield{author}{\bibinfo{person}{{Advanced Micro Devices, Inc.}}}
  \bibinfo{year}{2018}\natexlab{}.
\newblock \bibinfo{title}{{Software Techniques for Managing Speculation on AMD
  Processors, Revision 1.24.18}}.
\newblock   (\bibinfo{year}{2018}).
\newblock
\urldef\tempurl%
\url{https://developer.amd.com/wp-content/resources/Managing-Speculation-on-AMD-Processors.pdf}
\showURL{%
\tempurl}
\newblock
\shownote{[Online; accessed 09-June-2018].}


\bibitem[\protect\citeauthoryear{Akritidis, Cadar, Raiciu, Costa, and
  Castro}{Akritidis et~al\mbox{.}}{2008}]%
        {akritidis08sp-wit}
\bibfield{author}{\bibinfo{person}{Periklis Akritidis},
  \bibinfo{person}{Cristian Cadar}, \bibinfo{person}{Costin Raiciu},
  \bibinfo{person}{Manuel Costa}, {and} \bibinfo{person}{Miguel Castro}.}
  \bibinfo{year}{2008}\natexlab{}.
\newblock \showarticletitle{Preventing Memory Error Exploits with WIT}. In
  \bibinfo{booktitle}{\emph{Proceedings of the 2008 IEEE Symposium on Security
  and Privacy}} \emph{(\bibinfo{series}{SP '08})}. \bibinfo{publisher}{IEEE
  Computer Society}, \bibinfo{address}{Washington, DC, USA},
  \bibinfo{pages}{263--277}.
\newblock
\showISBNx{978-0-7695-3168-7}
\urldef\tempurl%
\url{https://doi.org/10.1109/SP.2008.30}
\showDOI{\tempurl}


\bibitem[\protect\citeauthoryear{ARM}{ARM}{2015}]%
        {arm15trm-cortexA72}
\bibfield{author}{\bibinfo{person}{ARM}.} \bibinfo{year}{2015}\natexlab{}.
\newblock \bibinfo{title}{{ARM} {Cortex-A72} {MPCore} Processor Technical
  Reference Manual}.
\newblock   (\bibinfo{year}{2015}).
\newblock


\bibitem[\protect\citeauthoryear{Boggs, Baktha, Hawkins, Marr, Miller, Roussel,
  Singhal, Toll, and Venkatraman}{Boggs et~al\mbox{.}}{2004}]%
        {boggs04itj-prescott}
\bibfield{author}{\bibinfo{person}{Darrell Boggs}, \bibinfo{person}{Aravindh
  Baktha}, \bibinfo{person}{Jason Hawkins}, \bibinfo{person}{Deborah Marr},
  \bibinfo{person}{J.~Alan Miller}, \bibinfo{person}{Patrice Roussel},
  \bibinfo{person}{Ronak Singhal}, \bibinfo{person}{Bret Toll}, {and}
  \bibinfo{person}{K.S. Venkatraman}.} \bibinfo{year}{2004}\natexlab{}.
\newblock \showarticletitle{The Microarchitecture of the {Intel Pentium 4}
  Processor on 90nm Technology.}
\newblock \bibinfo{journal}{\emph{Intel Technology Journal}}
  \bibinfo{volume}{8}, \bibinfo{number}{1} (\bibinfo{year}{2004}).
\newblock


\bibitem[\protect\citeauthoryear{Brasser, Davi, Gens, Liebchen, and
  Sadeghi}{Brasser et~al\mbox{.}}{2017}]%
        {brasser17sec-catt}
\bibfield{author}{\bibinfo{person}{Ferdinand Brasser}, \bibinfo{person}{Lucas
  Davi}, \bibinfo{person}{David Gens}, \bibinfo{person}{Christopher Liebchen},
  {and} \bibinfo{person}{Ahmad-Reza Sadeghi}.} \bibinfo{year}{2017}\natexlab{}.
\newblock \showarticletitle{{CA}n{\textquoteright}t {Touch This}: Software-only
  Mitigation against Rowhammer Attacks targeting Kernel Memory}. In
  \bibinfo{booktitle}{\emph{26th {USENIX} Security Symposium ({USENIX} Security
  17)}}. \bibinfo{publisher}{{USENIX} Association},
  \bibinfo{address}{Vancouver, BC}, \bibinfo{pages}{117--130}.
\newblock
\showISBNx{978-1-931971-40-9}
\urldef\tempurl%
\url{https://www.usenix.org/conference/usenixsecurity17/technical-sessions/presentation/brasser}
\showURL{%
\tempurl}


\bibitem[\protect\citeauthoryear{Brumley and Boneh}{Brumley and Boneh}{2005}]%
        {brumley2005rsa}
\bibfield{author}{\bibinfo{person}{David Brumley} {and} \bibinfo{person}{Dan
  Boneh}.} \bibinfo{year}{2005}\natexlab{}.
\newblock \showarticletitle{Remote timing attacks are practical}.
\newblock \bibinfo{journal}{\emph{Computer Networks}} (\bibinfo{year}{2005}).
\newblock


\bibitem[\protect\citeauthoryear{Carruth}{Carruth}{2018a}]%
        {llvm-spectre-variant-2}
\bibfield{author}{\bibinfo{person}{Chandler Carruth}.}
  \bibinfo{year}{2018}\natexlab{a}.
\newblock \bibinfo{title}{{Introduce the "retpoline" x86 mitigation technique
  for variant \#2 of the speculative execution vulnerabilities}}.
\newblock
  \bibinfo{howpublished}{\url{http://lists.llvm.org/pipermail/llvm-commits/Week-of-Mon-20180101/513630.html}}.
    (\bibinfo{date}{January} \bibinfo{year}{2018}).
\newblock


\bibitem[\protect\citeauthoryear{Carruth}{Carruth}{2018b}]%
        {llvm-spectre-variant-1}
\bibfield{author}{\bibinfo{person}{Chandler Carruth}.}
  \bibinfo{year}{2018}\natexlab{b}.
\newblock \bibinfo{title}{{Speculative Load Hardening: A Spectre Variant \#1
  Mitigation Technique}}.
\newblock   (\bibinfo{year}{2018}).
\newblock
\urldef\tempurl%
\url{https://lists.llvm.org/pipermail/llvm-dev/2018-March/122085.html}
\showURL{%
\tempurl}


\bibitem[\protect\citeauthoryear{Celio and Renau}{Celio and Renau}{2018}]%
        {ccelio18barcelona}
\bibfield{author}{\bibinfo{person}{Christopher Celio} {and}
  \bibinfo{person}{Jose Renau}.} \bibinfo{year}{2018}\natexlab{}.
\newblock \bibinfo{title}{{Securing High-performance {RISC-V} Processors from
  Time Speculation}}.
\newblock
  \bibinfo{howpublished}{\url{https://riscv.org/2018/05/risc-v-workshop-in-barcelona-proceedings/}}.
    (\bibinfo{date}{May} \bibinfo{year}{2018}).
\newblock


\bibitem[\protect\citeauthoryear{Chrysos and Emer}{Chrysos and Emer}{1998}]%
        {chrysos98isca-store}
\bibfield{author}{\bibinfo{person}{George~Z. Chrysos} {and}
  \bibinfo{person}{Joel~S. Emer}.} \bibinfo{year}{1998}\natexlab{}.
\newblock \showarticletitle{Memory Dependence Prediction Using Store Sets}. In
  \bibinfo{booktitle}{\emph{Proceedings of the 25th Annual International
  Symposium on Computer Architecture}} \emph{(\bibinfo{series}{ISCA '98})}.
  \bibinfo{address}{Washington, DC, USA}, \bibinfo{pages}{142--153}.
\newblock
\showISBNx{0-8186-8491-7}
\urldef\tempurl%
\url{https://doi.org/10.1145/279358.279378}
\showDOI{\tempurl}


\bibitem[\protect\citeauthoryear{{Common Weakness Enumeration}}{{Common
  Weakness Enumeration}}{2018}]%
        {cwe-cpp_weaknesses}
\bibfield{author}{\bibinfo{person}{{Common Weakness Enumeration}}.}
  \bibinfo{year}{2018}\natexlab{}.
\newblock \bibinfo{title}{Weaknesses in Software Written in C++}.
\newblock   (\bibinfo{year}{2018}).
\newblock
\urldef\tempurl%
\url{https://cwe.mitre.org/data/definitions/659.html}
\showURL{%
\tempurl}
\newblock
\shownote{[Online; accessed 09-June-2018].}


\bibitem[\protect\citeauthoryear{Corbet}{Corbet}{2017}]%
        {kaiser}
\bibfield{author}{\bibinfo{person}{Jonathan Corbet}.}
  \bibinfo{year}{2017}\natexlab{}.
\newblock \bibinfo{title}{{KAISER: hiding the kernel from user space}}.
\newblock \bibinfo{howpublished}{\url{https://lwn.net/Articles/738975/}}.
  (\bibinfo{date}{November} \bibinfo{year}{2017}).
\newblock


\bibitem[\protect\citeauthoryear{Evtyushkin, Ponomarev, and
  Abu-Ghazaleh}{Evtyushkin et~al\mbox{.}}{2016}]%
        {evtyushkin16micro-jaslr}
\bibfield{author}{\bibinfo{person}{Dmitry Evtyushkin}, \bibinfo{person}{Dmitry
  Ponomarev}, {and} \bibinfo{person}{Nael Abu-Ghazaleh}.}
  \bibinfo{year}{2016}\natexlab{}.
\newblock \showarticletitle{Jump over {ASLR}: Attacking Branch Predictors to
  Bypass {ASLR}}. In \bibinfo{booktitle}{\emph{The 49th Annual IEEE/ACM
  International Symposium on Microarchitecture}}
  \emph{(\bibinfo{series}{MICRO-49})}. \bibinfo{publisher}{IEEE Press},
  \bibinfo{address}{Piscataway, NJ, USA}, Article \bibinfo{articleno}{40},
  \bibinfo{numpages}{13}~pages.
\newblock
\urldef\tempurl%
\url{http://dl.acm.org/citation.cfm?id=3195638.3195686}
\showURL{%
\tempurl}


\bibitem[\protect\citeauthoryear{Frantzen and Shuey}{Frantzen and
  Shuey}{2001}]%
        {frantzen01sec-stackghost}
\bibfield{author}{\bibinfo{person}{Mike Frantzen} {and} \bibinfo{person}{Mike
  Shuey}.} \bibinfo{year}{2001}\natexlab{}.
\newblock \showarticletitle{{StackGhost}: Hardware Facilitated Stack
  Protection}. In \bibinfo{booktitle}{\emph{Proceedings of the 10th Conference
  on USENIX Security Symposium - Volume 10}}
  \emph{(\bibinfo{series}{SSYM'01})}. \bibinfo{publisher}{USENIX Association},
  \bibinfo{address}{Berkeley, CA, USA}, Article \bibinfo{articleno}{5}.
\newblock
\urldef\tempurl%
\url{http://dl.acm.org/citation.cfm?id=1251327.1251332}
\showURL{%
\tempurl}


\bibitem[\protect\citeauthoryear{Freier, Karlton, and Kocher}{Freier
  et~al\mbox{.}}{2011}]%
        {freier11rfc-ssl3}
\bibfield{author}{\bibinfo{person}{Alan~O. Freier}, \bibinfo{person}{Philip
  Karlton}, {and} \bibinfo{person}{Paul~C. Kocher}.}
  \bibinfo{year}{2011}\natexlab{}.
\newblock \showarticletitle{The Secure Sockets Layer {(SSL)} Protocol Version
  3.0}.
\newblock \bibinfo{journal}{\emph{{RFC}}}  \bibinfo{volume}{6101}
  (\bibinfo{year}{2011}), \bibinfo{pages}{1--67}.
\newblock
\urldef\tempurl%
\url{https://doi.org/10.17487/RFC6101}
\showDOI{\tempurl}


\bibitem[\protect\citeauthoryear{Ge, Yarom, Cock, and Heiser}{Ge
  et~al\mbox{.}}{2018}]%
        {ge2018survey}
\bibfield{author}{\bibinfo{person}{Qian Ge}, \bibinfo{person}{Yuval Yarom},
  \bibinfo{person}{David Cock}, {and} \bibinfo{person}{Gernot Heiser}.}
  \bibinfo{year}{2018}\natexlab{}.
\newblock \showarticletitle{A survey of microarchitectural timing attacks and
  countermeasures on contemporary hardware}.
\newblock \bibinfo{journal}{\emph{Journal of Cryptographic Engineering}}
  \bibinfo{volume}{8}, \bibinfo{number}{1} (\bibinfo{year}{2018}),
  \bibinfo{pages}{1--27}.
\newblock


\bibitem[\protect\citeauthoryear{Google}{Google}{2018}]%
        {google2018v8-65}
\bibfield{author}{\bibinfo{person}{Google}.} \bibinfo{year}{2018}\natexlab{}.
\newblock \bibinfo{title}{V8 {JavaScript} Engine}.
\newblock
  \bibinfo{howpublished}{\url{https://v8project.blogspot.com/2018/02/v8-release-65.html}}.
    (\bibinfo{date}{Feb} \bibinfo{year}{2018}).
\newblock


\bibitem[\protect\citeauthoryear{Gras and Razavi}{Gras and Razavi}{2017}]%
        {gras2017ndss-AnC}
\bibfield{author}{\bibinfo{person}{Ben Gras} {and} \bibinfo{person}{Kaveh
  Razavi}.} \bibinfo{year}{2017}\natexlab{}.
\newblock \showarticletitle{{ASLR} on the Line: Practical Cache Attacks on the
  {MMU}}.
\newblock \bibinfo{journal}{\emph{NDSS}} (\bibinfo{year}{2017}).
\newblock


\bibitem[\protect\citeauthoryear{Gruss, Maurice, and Mangard}{Gruss
  et~al\mbox{.}}{2016a}]%
        {gruss2016js-rowhammer}
\bibfield{author}{\bibinfo{person}{Daniel Gruss},
  \bibinfo{person}{Cl{\'e}mentine Maurice}, {and} \bibinfo{person}{Stefan
  Mangard}.} \bibinfo{year}{2016}\natexlab{a}.
\newblock \showarticletitle{Rowhammer.js: A remote software-induced fault
  attack in JavaScript}. In \bibinfo{booktitle}{\emph{International Conference
  on Detection of Intrusions and Malware, and Vulnerability Assessment}}.
  Springer, \bibinfo{pages}{300--321}.
\newblock


\bibitem[\protect\citeauthoryear{Gruss, Maurice, Wagner, and Mangard}{Gruss
  et~al\mbox{.}}{2016b}]%
        {gruss2016flushflush}
\bibfield{author}{\bibinfo{person}{Daniel Gruss},
  \bibinfo{person}{Cl{\'e}mentine Maurice}, \bibinfo{person}{Klaus Wagner},
  {and} \bibinfo{person}{Stefan Mangard}.} \bibinfo{year}{2016}\natexlab{b}.
\newblock \showarticletitle{{Flush+Flush}: a fast and stealthy cache attack}.
  In \bibinfo{booktitle}{\emph{International Conference on Detection of
  Intrusions and Malware, and Vulnerability Assessment}}. Springer,
  \bibinfo{pages}{279--299}.
\newblock


\bibitem[\protect\citeauthoryear{Horn}{Horn}{2018}]%
        {horn18spectre}
\bibfield{author}{\bibinfo{person}{Jann Horn}.}
  \bibinfo{year}{2018}\natexlab{}.
\newblock \bibinfo{title}{Reading privileged memory with a side-channel}.
\newblock
  \bibinfo{howpublished}{\url{https://googleprojectzero.blogspot.com/2018/01/}}.
    (\bibinfo{date}{January} \bibinfo{year}{2018}).
\newblock


\bibitem[\protect\citeauthoryear{{Intel}}{{Intel}}{2014}]%
        {intel14haswell-uncore}
\bibfield{author}{\bibinfo{person}{{Intel}}.} \bibinfo{year}{2014}\natexlab{}.
\newblock \bibinfo{title}{{Intel {Xeon Processor E5} v3 Family Uncore
  Performance Monitoring}}.
\newblock
  \bibinfo{howpublished}{\url{https://www-ssl.intel.com/content/www/us/en/processors/xeon/xeon-e5-v3-uncore-performance-monitoring.html}}.
    (\bibinfo{year}{2014}).
\newblock


\bibitem[\protect\citeauthoryear{{Intel}}{{Intel}}{2017a}]%
        {intel17cet}
\bibfield{author}{\bibinfo{person}{{Intel}}.} \bibinfo{year}{2017}\natexlab{a}.
\newblock \bibinfo{title}{{Control-flow Enforcement Technology Preview}}.
\newblock
  \bibinfo{howpublished}{\url{https://software.intel.com/sites/default/files/managed/4d/2a/control-flow-enforcement-technology-preview.pdf}}.
    (\bibinfo{year}{2017}).
\newblock


\bibitem[\protect\citeauthoryear{{Intel}}{{Intel}}{2017b}]%
        {intel17opt}
\bibfield{author}{\bibinfo{person}{{Intel}}.} \bibinfo{year}{2017}\natexlab{b}.
\newblock \bibinfo{title}{{Intel 64 and IA-32 Architectures Optimization
  Reference Manual}}.
\newblock
  \bibinfo{howpublished}{\url{http://www.intel.com/content/www/us/en/architecture-and-technology/64-ia-32-architectures-optimization-manual.html}}.
    (\bibinfo{year}{2017}).
\newblock
\newblock
\shownote{[Online; accessed 11-February-2018].}


\bibitem[\protect\citeauthoryear{{Intel}}{{Intel}}{2018a}]%
        {intel-ucode-spectre4}
\bibfield{author}{\bibinfo{person}{{Intel}}.} \bibinfo{year}{2018}\natexlab{a}.
\newblock \bibinfo{title}{Addressing New Research for Side-Channel Analysis:
  Details and Mitigation Information for Variant~4}.
\newblock   (\bibinfo{date}{May} \bibinfo{year}{2018}).
\newblock
\urldef\tempurl%
\url{https://newsroom.intel.com/editorials/addressing-new-research-for-side-channel-analysis/}
\showURL{%
\tempurl}
\newblock
\shownote{[Online; accessed 09-June-2018].}


\bibitem[\protect\citeauthoryear{{Intel}}{{Intel}}{2018b}]%
        {intel-bcb-whitepaper}
\bibfield{author}{\bibinfo{person}{{Intel}}.} \bibinfo{year}{2018}\natexlab{b}.
\newblock \bibinfo{title}{{Analyzing Potential Bounds Check Bypass
  Vulnerabilities}}.
\newblock   (\bibinfo{date}{July} \bibinfo{year}{2018}).
\newblock
\urldef\tempurl%
\url{https://software.intel.com/en-us/side-channel-security-support/}
\showURL{%
\tempurl}
\newblock
\shownote{[Online; to appear 10-July-2018].}


\bibitem[\protect\citeauthoryear{{Intel}}{{Intel}}{2018c}]%
        {intel-sdm-vol3}
\bibfield{author}{\bibinfo{person}{{Intel}}.} \bibinfo{year}{2018}\natexlab{c}.
\newblock \bibinfo{title}{{Intel 64 and IA-32 Architectures Software
  Developer's Manual, Volume 3: System Programming Guide}}.
\newblock   (\bibinfo{date}{May} \bibinfo{year}{2018}).
\newblock
\urldef\tempurl%
\url{https://www.intel.com/sdm/}
\showURL{%
\tempurl}


\bibitem[\protect\citeauthoryear{{Intel}}{{Intel}}{2018d}]%
        {intel18spec-v4}
\bibfield{author}{\bibinfo{person}{{Intel}}.} \bibinfo{year}{2018}\natexlab{d}.
\newblock \bibinfo{title}{{Speculative Execution Side Channel Mitigations.
  Revision 2.0}}.
\newblock
  \bibinfo{howpublished}{\url{https://software.intel.com/sites/default/files/managed/c5/63/336996-Speculative-Execution-Side-Channel-Mitigations.pdf}}.
    (\bibinfo{date}{May} \bibinfo{year}{2018}).
\newblock


\bibitem[\protect\citeauthoryear{Kim, Daly, Kim, Fallin, Lee, Lee, Wilkerson,
  Lai, and Mutlu}{Kim et~al\mbox{.}}{2014}]%
        {kim2014rowhammer}
\bibfield{author}{\bibinfo{person}{Yoongu Kim}, \bibinfo{person}{Ross Daly},
  \bibinfo{person}{Jeremie Kim}, \bibinfo{person}{Chris Fallin},
  \bibinfo{person}{Ji~Hye Lee}, \bibinfo{person}{Donghyuk Lee},
  \bibinfo{person}{Chris Wilkerson}, \bibinfo{person}{Konrad Lai}, {and}
  \bibinfo{person}{Onur Mutlu}.} \bibinfo{year}{2014}\natexlab{}.
\newblock \showarticletitle{Flipping bits in memory without accessing them: An
  experimental study of {DRAM} disturbance errors}. In
  \bibinfo{booktitle}{\emph{ISCA}}. IEEE Press.
\newblock


\bibitem[\protect\citeauthoryear{Kiriansky}{Kiriansky}{2003}]%
        {kiriansky03meng-rio}
\bibfield{author}{\bibinfo{person}{Vladimir Kiriansky}.}
  \bibinfo{year}{2003}\natexlab{}.
\newblock \emph{\bibinfo{title}{Secure Execution Environment via Program
  Shepherding}}.
\newblock M.Eng. Thesis. \bibinfo{school}{Massachusetts Institute of
  Technology}, \bibinfo{address}{Cambridge, MA}.
\newblock
\urldef\tempurl%
\url{http://groups.csail.mit.edu/commit/papers/03/vlk-MEthesis.pdf}
\showURL{%
\tempurl}


\bibitem[\protect\citeauthoryear{Kiriansky, Bruening, and
  Amarasinghe}{Kiriansky et~al\mbox{.}}{2002}]%
        {kiriansky02sec-shepherding}
\bibfield{author}{\bibinfo{person}{Vladimir Kiriansky}, \bibinfo{person}{Derek
  Bruening}, {and} \bibinfo{person}{Saman Amarasinghe}.}
  \bibinfo{year}{2002}\natexlab{}.
\newblock \showarticletitle{Secure Execution via Program Shepherding}. In
  \bibinfo{booktitle}{\emph{Proceedings of the 11th USENIX Security
  Symposium}}. \bibinfo{publisher}{USENIX Association},
  \bibinfo{address}{Berkeley, CA, USA}, \bibinfo{pages}{191--206}.
\newblock
\showISBNx{1-931971-00-5}
\urldef\tempurl%
\url{http://dl.acm.org/citation.cfm?id=647253.720293}
\showURL{%
\tempurl}


\bibitem[\protect\citeauthoryear{Kiriansky, Bruening, and
  Amarasinghe}{Kiriansky et~al\mbox{.}}{2003}]%
        {kiriansky03tm-cfi-pldi-rejection}
\bibfield{author}{\bibinfo{person}{Vladimir Kiriansky}, \bibinfo{person}{Derek
  Bruening}, {and} \bibinfo{person}{Saman Amarasinghe}.}
  \bibinfo{year}{2003}\natexlab{}.
\newblock \bibinfo{booktitle}{\emph{Execution Model Enforcement Via Program
  Shepherding}}.
\newblock \bibinfo{type}{Technical Report} MIT/LCS Technical Memo LCS-TM-638.
  \bibinfo{institution}{Massachusetts Institute of Technology},
  \bibinfo{address}{Cambridge, MA}.
\newblock
\urldef\tempurl%
\url{http://groups.csail.mit.edu/commit/papers/03/RIO-security-TM-638.pdf}
\showURL{%
\tempurl}


\bibitem[\protect\citeauthoryear{Kiriansky, Lebedev, Amarasinghe, Devadas, and
  Emer}{Kiriansky et~al\mbox{.}}{2018}]%
        {kiriansky18crypto-dawg}
\bibfield{author}{\bibinfo{person}{Vladimir Kiriansky}, \bibinfo{person}{Ilia
  Lebedev}, \bibinfo{person}{Saman Amarasinghe}, \bibinfo{person}{Srinivas
  Devadas}, {and} \bibinfo{person}{Joel Emer}.}
  \bibinfo{year}{2018}\natexlab{}.
\newblock \bibinfo{title}{{DAWG}: A Defense Against Cache Timing Attacks in
  Speculative Execution Processors}.
\newblock \bibinfo{howpublished}{Cryptology ePrint Archive, Report 2018/418}.
  (\bibinfo{date}{May} \bibinfo{year}{2018}).
\newblock
\urldef\tempurl%
\url{https://eprint.iacr.org/2018/418}
\showURL{%
\tempurl}


\bibitem[\protect\citeauthoryear{Kocher, Genkin, Gruss, Haas, Hamburg, Lipp,
  Mangard, Prescher, Schwarz, and Yarom}{Kocher et~al\mbox{.}}{2018}]%
        {Kocher2018spectre}
\bibfield{author}{\bibinfo{person}{Paul Kocher}, \bibinfo{person}{Daniel
  Genkin}, \bibinfo{person}{Daniel Gruss}, \bibinfo{person}{Werner Haas},
  \bibinfo{person}{Mike Hamburg}, \bibinfo{person}{Moritz Lipp},
  \bibinfo{person}{Stefan Mangard}, \bibinfo{person}{Thomas Prescher},
  \bibinfo{person}{Michael Schwarz}, {and} \bibinfo{person}{Yuval Yarom}.}
  \bibinfo{year}{2018}\natexlab{}.
\newblock \showarticletitle{Spectre Attacks: Exploiting Speculative Execution}.
\newblock \bibinfo{journal}{\emph{ArXiv e-prints}} (\bibinfo{date}{Jan.}
  \bibinfo{year}{2018}).
\newblock
\showeprint[arxiv]{1801.01203}


\bibitem[\protect\citeauthoryear{Kroft}{Kroft}{1981}]%
        {kroft81isca-mshr}
\bibfield{author}{\bibinfo{person}{David Kroft}.}
  \bibinfo{year}{1981}\natexlab{}.
\newblock \showarticletitle{Lockup-free Instruction Fetch/Prefetch Cache
  Organization}. In \bibinfo{booktitle}{\emph{Proceedings of the 8th Annual
  Symposium on Computer Architecture}} \emph{(\bibinfo{series}{ISCA '81})}.
  \bibinfo{publisher}{IEEE Computer Society Press}, \bibinfo{address}{Los
  Alamitos, CA, USA}, \bibinfo{pages}{81--87}.
\newblock
\urldef\tempurl%
\url{http://dl.acm.org/citation.cfm?id=800052.801868}
\showURL{%
\tempurl}


\bibitem[\protect\citeauthoryear{Kuznetsov, Szekeres, Payer, Candea, Sekar, and
  Song}{Kuznetsov et~al\mbox{.}}{2014}]%
        {kuznetsov14osdi-cpi}
\bibfield{author}{\bibinfo{person}{Volodymyr Kuznetsov},
  \bibinfo{person}{L\'{a}szl\'{o} Szekeres}, \bibinfo{person}{Mathias Payer},
  \bibinfo{person}{George Candea}, \bibinfo{person}{R. Sekar}, {and}
  \bibinfo{person}{Dawn Song}.} \bibinfo{year}{2014}\natexlab{}.
\newblock \showarticletitle{Code-pointer Integrity}. In
  \bibinfo{booktitle}{\emph{Proceedings of the 11th USENIX Conference on
  Operating Systems Design and Implementation}}
  \emph{(\bibinfo{series}{OSDI'14})}. \bibinfo{publisher}{USENIX Association},
  \bibinfo{address}{Berkeley, CA, USA}, \bibinfo{pages}{147--163}.
\newblock
\showISBNx{978-1-931971-16-4}
\urldef\tempurl%
\url{http://dl.acm.org/citation.cfm?id=2685048.2685061}
\showURL{%
\tempurl}


\bibitem[\protect\citeauthoryear{{Linux Kernel}}{{Linux Kernel}}{2018}]%
        {linux-doc-speculation}
\bibfield{author}{\bibinfo{person}{{Linux Kernel}}.}
  \bibinfo{year}{2018}\natexlab{}.
\newblock \bibinfo{title}{{Speculation Documentation}}.
\newblock   (\bibinfo{year}{2018}).
\newblock
\urldef\tempurl%
\url{https://www.kernel.org/doc/Documentation/speculation.txt}
\showURL{%
\tempurl}
\newblock
\shownote{[Online; accessed 09-June-2018].}


\bibitem[\protect\citeauthoryear{Lipp, Schwarz, Gruss, Prescher, Haas, Mangard,
  Kocher, Genkin, Yarom, and Hamburg}{Lipp et~al\mbox{.}}{2018}]%
        {Lipp2018meltdown}
\bibfield{author}{\bibinfo{person}{Moritz Lipp}, \bibinfo{person}{Michael
  Schwarz}, \bibinfo{person}{Daniel Gruss}, \bibinfo{person}{Thomas Prescher},
  \bibinfo{person}{Werner Haas}, \bibinfo{person}{Stefan Mangard},
  \bibinfo{person}{Paul Kocher}, \bibinfo{person}{Daniel Genkin},
  \bibinfo{person}{Yuval Yarom}, {and} \bibinfo{person}{Mike Hamburg}.}
  \bibinfo{year}{2018}\natexlab{}.
\newblock \showarticletitle{Meltdown}.
\newblock \bibinfo{journal}{\emph{ArXiv e-prints}} (\bibinfo{date}{Jan.}
  \bibinfo{year}{2018}).
\newblock
\showeprint[arxiv]{1801.01207}


\bibitem[\protect\citeauthoryear{Liu, Yarom, Ge, Heiser, and Lee}{Liu
  et~al\mbox{.}}{2015}]%
        {liu2015llctiming}
\bibfield{author}{\bibinfo{person}{Fangfei Liu}, \bibinfo{person}{Yuval Yarom},
  \bibinfo{person}{Qian Ge}, \bibinfo{person}{Gernot Heiser}, {and}
  \bibinfo{person}{Ruby~B Lee}.} \bibinfo{year}{2015}\natexlab{}.
\newblock \showarticletitle{Last-Level Cache Side-Channel Attacks are
  Practical}. In \bibinfo{booktitle}{\emph{Security and Privacy}}. IEEE.
\newblock


\bibitem[\protect\citeauthoryear{Lu, Walter, Pfaff, N{\"u}rnberger, Lee, and
  Backes}{Lu et~al\mbox{.}}{2017}]%
        {lu2017ndss-stackspray}
\bibfield{author}{\bibinfo{person}{Kangjie Lu}, \bibinfo{person}{Marie-Therese
  Walter}, \bibinfo{person}{David Pfaff}, \bibinfo{person}{Stefan
  N{\"u}rnberger}, \bibinfo{person}{Wenke Lee}, {and} \bibinfo{person}{Michael
  Backes}.} \bibinfo{year}{2017}\natexlab{}.
\newblock \showarticletitle{Unleashing use-before-initialization
  vulnerabilities in the Linux kernel using targeted stack spraying}
  \emph{(\bibinfo{series}{NDSS'17})}.
\newblock


\bibitem[\protect\citeauthoryear{Microsoft}{Microsoft}{2013}]%
        {microsoft13eaten-cookie}
\bibfield{author}{\bibinfo{person}{Microsoft}.}
  \bibinfo{year}{2013}\natexlab{}.
\newblock \bibinfo{title}{{Software defense: mitigating stack corruption
  vulnerabilties}}.
\newblock
  \bibinfo{howpublished}{\url{https://blogs.technet.microsoft.com/srd/2013/10/02/software-defense-mitigating-stack-corruption-vulnerabilties/}}.
    (\bibinfo{year}{2013}).
\newblock


\bibitem[\protect\citeauthoryear{{Microsoft}}{{Microsoft}}{2018}]%
        {microsoft-chakra-spectre}
\bibfield{author}{\bibinfo{person}{{Microsoft}}.}
  \bibinfo{year}{2018}\natexlab{}.
\newblock \bibinfo{title}{{Add JIT mitigations for Spectre}}.
\newblock   (\bibinfo{date}{February} \bibinfo{year}{2018}).
\newblock
\urldef\tempurl%
\url{https://github.com/Microsoft/ChakraCore/commit/08b82b8d33e9b36c0d6628b856f280234c87ba13}
\showURL{%
\tempurl}


\bibitem[\protect\citeauthoryear{{Mozilla}}{{Mozilla}}{2018}]%
        {performance-now}
\bibfield{author}{\bibinfo{person}{{Mozilla}}.}
  \bibinfo{year}{2018}\natexlab{}.
\newblock \bibinfo{title}{{MDN web docs} --- performance.now()}.
\newblock   (\bibinfo{year}{2018}).
\newblock
\urldef\tempurl%
\url{https://developer.mozilla.org/en-US/docs/Web/API/Performance/now}
\showURL{%
\tempurl}
\newblock
\shownote{[Online; accessed 09-June-2018].}


\bibitem[\protect\citeauthoryear{Oren, Kemerlis, Sethumadhavan, and
  Keromytis}{Oren et~al\mbox{.}}{2015}]%
        {oren15ccs-spy}
\bibfield{author}{\bibinfo{person}{Yossef Oren}, \bibinfo{person}{Vasileios~P.
  Kemerlis}, \bibinfo{person}{Simha Sethumadhavan}, {and}
  \bibinfo{person}{Angelos~D. Keromytis}.} \bibinfo{year}{2015}\natexlab{}.
\newblock \showarticletitle{The Spy in the Sandbox: Practical Cache Attacks in
  JavaScript and Their Implications}. In \bibinfo{booktitle}{\emph{Proceedings
  of the 22nd ACM SIGSAC Conference on Computer and Communications Security}}
  \emph{(\bibinfo{series}{CCS '15})}. \bibinfo{publisher}{ACM},
  \bibinfo{address}{New York, NY, USA}, \bibinfo{pages}{1406--1418}.
\newblock
\showISBNx{978-1-4503-3832-5}
\urldef\tempurl%
\url{https://doi.org/10.1145/2810103.2813708}
\showDOI{\tempurl}


\bibitem[\protect\citeauthoryear{Pardoe}{Pardoe}{2018}]%
        {msvc-spectre-variant-1}
\bibfield{author}{\bibinfo{person}{Andrew Pardoe}.}
  \bibinfo{year}{2018}\natexlab{}.
\newblock \bibinfo{title}{{Spectre mitigations in MSVC}}.
\newblock
  \bibinfo{howpublished}{\url{https://blogs.msdn.microsoft.com/vcblog/2018/01/15/spectre-mitigations-in-msvc/}}.
    (\bibinfo{date}{January} \bibinfo{year}{2018}).
\newblock


\bibitem[\protect\citeauthoryear{{Richard Grisenthwaite}}{{Richard
  Grisenthwaite}}{2018}]%
        {grisenthwaite18v1}
\bibfield{author}{\bibinfo{person}{{Richard Grisenthwaite}}.}
  \bibinfo{year}{2018}\natexlab{}.
\newblock \bibinfo{title}{{Cache Speculation Side-channels}}.
\newblock
  \bibinfo{howpublished}{\url{https://developer.arm.com/support/arm-security-updates/speculative-processor-vulnerability}}.
    (\bibinfo{date}{January} \bibinfo{year}{2018}).
\newblock


\bibitem[\protect\citeauthoryear{Seaborn and Dullien}{Seaborn and
  Dullien}{2015}]%
        {google2015rowhammer}
\bibfield{author}{\bibinfo{person}{Mark Seaborn} {and} \bibinfo{person}{Thomas
  Dullien}.} \bibinfo{year}{2015}\natexlab{}.
\newblock \bibinfo{title}{Exploiting the {DRAM} {RowHammer} bug to gain kernel
  privileges}.
\newblock
  \bibinfo{howpublished}{\url{http://googleprojectzero.blogspot.com/2015/03/exploiting-dram-rowhammer-bug-to-gain.html}}.
    (\bibinfo{date}{Mar} \bibinfo{year}{2015}).
\newblock
\newblock
\shownote{[Online; accessed 19-February-2018].}


\bibitem[\protect\citeauthoryear{Shankland}{Shankland}{2005}]%
        {cnet2005itanium}
\bibfield{author}{\bibinfo{person}{Stephen Shankland}.}
  \bibinfo{year}{2005}\natexlab{}.
\newblock \showarticletitle{Itanium: A cautionary tale}.
\newblock
  \bibinfo{howpublished}{\url{http://news.cnet.com/Itanium-A-cautionary-tale/2100-1006_3-5984747.html}}.
\newblock  (\bibinfo{date}{Dec} \bibinfo{year}{2005}).
\newblock
\newblock
\shownote{[Online; accessed 15-January-2018].}


\bibitem[\protect\citeauthoryear{Sidiroglou{-}Douskos}{Sidiroglou{-}Douskos}{2018}]%
        {spechammer}
\bibfield{author}{\bibinfo{person}{Stelios Sidiroglou{-}Douskos}.}
  \bibinfo{year}{2018}\natexlab{}.
\newblock \bibinfo{title}{{SpecHammer = Spectre1.1 + RowHammer}}.
\newblock \bibinfo{howpublished}{Private communication}.
  (\bibinfo{date}{February} \bibinfo{year}{2018}).
\newblock


\bibitem[\protect\citeauthoryear{Tatar, Konoth, Athanasopoulos, Giuffrida, Bos,
  and Razavi}{Tatar et~al\mbox{.}}{2018}]%
        {tatar18usenix-throwhammer}
\bibfield{author}{\bibinfo{person}{Andrei Tatar},
  \bibinfo{person}{Radhesh~Krishnan Konoth}, \bibinfo{person}{Elias
  Athanasopoulos}, \bibinfo{person}{Cristiano Giuffrida},
  \bibinfo{person}{Herbert Bos}, {and} \bibinfo{person}{Kaveh Razavi}.}
  \bibinfo{year}{2018}\natexlab{}.
\newblock \showarticletitle{Throwhammer: Rowhammer Attacks over the Network and
  Defenses}. In \bibinfo{booktitle}{\emph{2018 {USENIX} Annual Technical
  Conference ({USENIX} {ATC} 18)}}. \bibinfo{publisher}{{USENIX} Association},
  \bibinfo{address}{Boston, MA}.
\newblock
\urldef\tempurl%
\url{https://www.usenix.org/conference/atc18/presentation/tatar}
\showURL{%
\tempurl}


\bibitem[\protect\citeauthoryear{Tice, Roeder, Collingbourne, Checkoway,
  Erlingsson, Lozano, and Pike}{Tice et~al\mbox{.}}{2014}]%
        {tice14sec-cfigcc}
\bibfield{author}{\bibinfo{person}{Caroline Tice}, \bibinfo{person}{Tom
  Roeder}, \bibinfo{person}{Peter Collingbourne}, \bibinfo{person}{Stephen
  Checkoway}, \bibinfo{person}{\'{U}lfar Erlingsson}, \bibinfo{person}{Luis
  Lozano}, {and} \bibinfo{person}{Geoff Pike}.}
  \bibinfo{year}{2014}\natexlab{}.
\newblock \showarticletitle{Enforcing Forward-edge Control-flow Integrity in
  GCC \& LLVM}. In \bibinfo{booktitle}{\emph{Proceedings of the 23rd USENIX
  Conference on Security Symposium}} \emph{(\bibinfo{series}{SEC'14})}.
  \bibinfo{publisher}{USENIX Association}, \bibinfo{address}{Berkeley, CA,
  USA}, \bibinfo{pages}{941--955}.
\newblock
\showISBNx{978-1-931971-15-7}
\urldef\tempurl%
\url{http://dl.acm.org/citation.cfm?id=2671225.2671285}
\showURL{%
\tempurl}


\bibitem[\protect\citeauthoryear{Turner}{Turner}{2018}]%
        {retpoline}
\bibfield{author}{\bibinfo{person}{Paul Turner}.}
  \bibinfo{year}{2018}\natexlab{}.
\newblock \bibinfo{title}{{Retpoline: a software construct for preventing
  branch-target-injection}}.
\newblock
  \bibinfo{howpublished}{\url{https://support.google.com/faqs/answer/7625886}}.
    (\bibinfo{date}{January} \bibinfo{year}{2018}).
\newblock


\bibitem[\protect\citeauthoryear{Vahldiek{-}Oberwagner, Elnikety, Garg, and
  Druschel}{Vahldiek{-}Oberwagner et~al\mbox{.}}{2018}]%
        {vahldiek18corr-erim}
\bibfield{author}{\bibinfo{person}{Anjo Vahldiek{-}Oberwagner},
  \bibinfo{person}{Eslam Elnikety}, \bibinfo{person}{Deepak Garg}, {and}
  \bibinfo{person}{Peter Druschel}.} \bibinfo{year}{2018}\natexlab{}.
\newblock \showarticletitle{{ERIM:} Secure and Efficient In-process Isolation
  with Memory Protection Keys}.
\newblock \bibinfo{journal}{\emph{CoRR}}  \bibinfo{volume}{abs/1801.06822}
  (\bibinfo{year}{2018}).
\newblock
\showeprint[arxiv]{1801.06822}
\urldef\tempurl%
\url{http://arxiv.org/abs/1801.06822}
\showURL{%
\tempurl}


\bibitem[\protect\citeauthoryear{Waterman, Lee, Patterson, and
  Asanovi\'{c}}{Waterman et~al\mbox{.}}{2014}]%
        {waterman2014riscv}
\bibfield{author}{\bibinfo{person}{Andrew Waterman}, \bibinfo{person}{Yunsup
  Lee}, \bibinfo{person}{David~A. Patterson}, {and} \bibinfo{person}{Krste
  Asanovi\'{c}}.} \bibinfo{year}{2014}\natexlab{}.
\newblock \bibinfo{booktitle}{\emph{The {RISC-V} Instruction Set Manual, Volume
  I: User-Level {ISA}, Version 2.0}}.
\newblock \bibinfo{type}{{T}echnical {R}eport} UCB/EECS-2014-54.
  \bibinfo{institution}{EECS Department, University of California, Berkeley}.
\newblock
\urldef\tempurl%
\url{http://www.eecs.berkeley.edu/Pubs/TechRpts/2014/EECS-2014-54.html}
\showURL{%
\tempurl}


\bibitem[\protect\citeauthoryear{Weston and Miller}{Weston and Miller}{2016}]%
        {weston16blackhat-cfg}
\bibfield{author}{\bibinfo{person}{David Weston} {and} \bibinfo{person}{Matt
  Miller}.} \bibinfo{year}{2016}\natexlab{}.
\newblock \bibinfo{title}{{Windows 10 Mitigation Improvements}}.
\newblock
  \bibinfo{howpublished}{\url{https://www.blackhat.com/docs/us-16/materials/us-16-Weston-Windows-10-Mitigation-Improvements.pdf}}.
    (\bibinfo{year}{2016}).
\newblock


\bibitem[\protect\citeauthoryear{{Wikipedia}}{{Wikipedia}}{2018a}]%
        {wiki-eternalblue}
\bibfield{author}{\bibinfo{person}{{Wikipedia}}.}
  \bibinfo{year}{2018}\natexlab{a}.
\newblock \bibinfo{title}{{EternalBlue}}.
\newblock   (\bibinfo{year}{2018}).
\newblock
\urldef\tempurl%
\url{https://en.wikipedia.org/wiki/EternalBlue}
\showURL{%
\tempurl}
\newblock
\shownote{[Online; accessed 09-June-2018].}


\bibitem[\protect\citeauthoryear{{Wikipedia}}{{Wikipedia}}{2018b}]%
        {wiki-memory_safety}
\bibfield{author}{\bibinfo{person}{{Wikipedia}}.}
  \bibinfo{year}{2018}\natexlab{b}.
\newblock \bibinfo{title}{{Memory safety} --- Types of memory errors}.
\newblock   (\bibinfo{year}{2018}).
\newblock
\urldef\tempurl%
\url{https://en.wikipedia.org/wiki/Memory_safety#Types_of_memory_errors}
\showURL{%
\tempurl}
\newblock
\shownote{[Online; accessed 09-June-2018].}


\bibitem[\protect\citeauthoryear{Yarom and Falkner}{Yarom and Falkner}{2014}]%
        {yarom2014flush}
\bibfield{author}{\bibinfo{person}{Yuval Yarom} {and} \bibinfo{person}{Katrina
  Falkner}.} \bibinfo{year}{2014}\natexlab{}.
\newblock \showarticletitle{{FLUSH+RELOAD}: A High Resolution, Low Noise, {L3}
  Cache Side-Channel Attack}. In \bibinfo{booktitle}{\emph{USENIX Security
  Symposium}}.
\newblock


\bibitem[\protect\citeauthoryear{Yeager}{Yeager}{1996}]%
        {yeager96mipsr10k}
\bibfield{author}{\bibinfo{person}{K.C. Yeager}.}
  \bibinfo{year}{1996}\natexlab{}.
\newblock \showarticletitle{{The MIPS R10000 Superscalar Microprocessor}}.
\newblock \bibinfo{journal}{\emph{IEEE Micro}} (\bibinfo{year}{1996}).
\newblock
\showISSN{0272-1732}
\urldef\tempurl%
\url{https://doi.org/10.1109/40.491460}
\showDOI{\tempurl}


\end{thebibliography}

\end{document}